\documentclass[printer]{aa}

\usepackage{graphicx}
\usepackage{subcaption}
\usepackage{amssymb}
\usepackage{amsmath}
\usepackage{float}
\usepackage{txfonts}
\usepackage{gensymb}
\usepackage{hyperref}
\usepackage[normalem]{ulem}
\usepackage[figuresright]{rotating}

\usepackage{amsmath}
\usepackage{threeparttable}
\usepackage{footnote}
\usepackage{color}
\usepackage[figuresright]{rotating}

\def\simgt{\lower 2pt \hbox{$\, \buildrel {\scriptstyle >}\over {\scriptstyle \sim}\,$}}
\def\simlt{\lower 2pt \hbox{$\, \buildrel {\scriptstyle <}\over {\scriptstyle \sim}\,$}}

\authorrunning{Junhui Liu}
\titlerunning{X-ray Emission of Contact Binaries}

\hypersetup{draft}
\begin{document}

\title{X-ray emission of contact binary variables within 1~kpc \footnote{The data in Tables \ref{table:Properties_EWXs}, \ref{table:SRASS_EWXs}, \ref{table:Spectra_EWXs}, \ref{table:XMMnewtonupperlimits} and \ref{table:ROSATupperlimits} are only available at the CDS via anonymous ftp to cdsarc.u-strasbg.fr (130.79.128.5) or via http://cdsarc.u-strasbg.fr/viz-bin/cat/J/A+A/}}

\author{Junhui Liu \inst{1,2,3} \and Jianfeng Wu \inst{1} \and Ali Esamdin \inst{2,3} \and Wei-Min Gu \inst{1} \and Mouyuan Sun \inst{1} \and Junfeng Wang \inst{1}}

\institute{ Department of Astronomy, Xiamen University, Xiamen, Fujian 361005, People's Republic of China (\email{wujianfeng@xmu.edu.cn}\label{inst1}) 
\and
 Xinjiang Astronomical Observatory, Chinese Academy of Sciences, Urumqi, Xinjiang 830011, People's Republic of China (\email{aliyi@xao.ac.cn})
\and
University of Chinese Academy of Sciences, Beijing 100049, People's Republic of China\label{inst2}}

\date{Received - - - ; accepted - - - }

\abstract{}{The X-ray emission of contact binaries (EW-type) is an important facet of such systems. Thus, X-ray emitting EW-type binaries (EWXs) are ideal laboratories for  studying the X-ray radiation saturation mechanisms as well as binary evolution. By assembling the largest sample to date of EWXs with periods of less than 1 day from the All-Sky Automated Survey for Supernovae Variable Stars Database and X-ray catalogs from the XMM-Newton and ROSAT missions, we aim to conduct a systematic population study of X-ray emission properties of EWXs within 1 kpc.}
{We carried out correlation analyses for the X-ray luminosity, log$L_{\textrm{X}}$, and X-ray activity level log($L_{\textrm{X}}$/$L_{\textrm{bol}}$) versus the orbital period, $P$, effective temperature, $T_{\rm eff}$, metallicity [Fe/H], and the surface gravity $\log g$ of EWXs. We investigated the relation between X-ray emission and the mass of component stars in the binary systems. We also performed sample simulations to explore the degeneracy between period, mass, and effective temperature for EWXs.}
{We find strong $P$-log$L_{\textrm{X}}$ and $P$-log($L_{\textrm{X}}$/$L_{\textrm{bol}}$) correlations for EWXs with $P\simlt0.44$ days and we provide the linear parametrizations for these relations, on the basis of which the orbital period can be treated as a good predictor for log$L_{\textrm{X}}$ and log($L_{\textrm{X}}$/$L_{\textrm{bol}}$). The aforementioned binary stellar parameters are all correlated with log$L_{\textrm{X}}$, while only $T_{\rm eff}$ exhibits a strong correlation with log($L_{\textrm{X}}$/$L_{\textrm{bol}}$). Then, EWXs with higher temperature show lower X-ray activity level, which could indicate the thinning of the convective area related to the magnetic dynamo mechanism. The total X-ray luminosity of an EWX is essentially consistent with that of an X-ray saturated main sequence star with the same mass as its primary, which may imply that the primary star dominates the X-ray emission. The monotonically decreasing $P$-log($L_{\textrm{X}}$/$L_{\textrm{bol}}$) relation and the short orbital periods indicate that EWXs could all be in the X-ray saturated state, and they may inherit the changing trend of the saturated X-ray luminosities along with the mass shown by single stars. For EWXs, the orbital period, mass, and effective temperature increase in concordance. We demonstrate that the period $P=0.44$ days corresponds to the primary mass of $\sim1.1 \rm M_\odot$, beyond which the saturated X-ray luminosity of single stars will not continue to increase with mass. This explains the break in the positive $P$-log$L_{\textrm{X}}$ relation for EWXs with $P>0.44$ days.}{}

\maketitle{}

\section{Introduction} \label{sec:intro}

\par

The W Ursa Majoris (W~UMa)-type binary, also known as an EW-type binary, is a contact system where the two components fill their Roche lobes and share a common envelope. 
The spectral types of EW binaries usually range from F to K \citep{2017RAA....17...87Q}.
Their variability amplitudes are generally less than 1 magnitude, while the typical orbital period ranges from 0.2 to 1.0 day, but strongly
peaked between 0.2-0.5 days \citep[see][Figure~1]{2017RAA....17...87Q}.
The iconic feature of their optical light curves is the continuous variation of luminosity with nearly equal depths of the
primary and secondary minima, which indicates that the two components are in thermal contact, characterized by nearly identical temperatures. Furthermore, EWs can be subsequently divided into two subtypes based on mass and temperature, namely: the W subtype and A subtype. The primary star of the former is a more-massive and hotter component, while that of the latter is a more-massive and cooler one.
A fraction of W~UMa systems are X-ray sources, which we refer to as EW-type binaries with X-ray emission (hereafter EWXs).
Although the X-ray emission mechanism remains puzzling, the stellar dynamo magnetic activity generated by rapid rotation and envelope convection is usually in consideration \citep{2004A&A...415.1113G}.
The studies of BH~Cas \citep{2019PASP..131h4202L} and 2MASS~J11201034$-$2201340 \citep{2016AJ....151..170H} indicate that the X-ray light curves do not show any obvious occultation or modulation as optical light curves would. The X-ray spectra of the former can be described by thermal models, while those of the latter can be both fitted by a thermal or a power-law model. The X-ray grating spectra of VW~Cep \citep{2006ApJ...650.1119H} reveal that the compact corona is mainly located in the polar region of the primary star.

\par

\citet{2001A&A...370..157S} examined a sample of 102 W~UMa systems and found 57 of them were X-ray sources detected by the ROSAT All Sky Survey (RASS), which indicates a high fraction of W~UMa binaries having X-ray emission. 
\citet{2006AJ....131..990C} also obtained X-ray fluxes for 34 W~UMa  systems from the RASS database. In these studies, they calculated the hardness ratio between the X-ray counts in the hard (H; 0.5-2.0 keV) and soft (S; 0.1-0.4 keV) bands, and then the X-ray flux and X-ray luminosity based on the energy conversion factor derived from hardness ratio \citep{1996A&A...310..801H}. The X-ray luminosities of W~UMa type binaries within 400 pc range from $4.4 \times 10^{29}$ to $2.3 \times 10^{31}$ erg s$^{-1}$ \citep{2006AJ....131..990C}. Combining the samples of contact binaries observed by ROTSE-1 and the RASS catalog, \citet{2006AJ....131..633G}  found that 140 contact binaries have X-ray emission with typical luminosities of $\sim 1.0 \times 10^{30}$erg s$^{-1}$. \citet{2008AcA....58..405S} compiled a catalog containing 379 X-ray emitting contact eclipsing binaries for which they applied somewhat different selection criteria from the widely-used EW classification. They found evidence of an X-ray saturation effect, while their sample exhibits large scatter in the X-ray activity and period relation.

\par

For late-type main sequence stars (G- to F-types), their X-ray luminosities, $L_{\textrm{X}}$, tend to reach a maximum value at 10$^{-3}$ of the star's bolometric luminosity $L_{\textrm{bol}}$, namely, log($L_{\textrm{X}}$/$L_{\textrm{bol}}$) $\sim -3$, which is known as the "saturation limit," while the $\log(L_{\textrm{X}}$/$L_{\textrm{bol}}$) can be used to describe the X-ray activity level \citep{1984A&A...133..117V, 1987ApJ...321..958V, 1993ApJ...410..387F}. For single stars with $P<$~0.4~days, the phenomenon of log($L_{\textrm{X}}$/$L_{\textrm{bol}}$) decreasing with the shortening period is referred to as "supersaturation"\ \citep{1984ApJ...277..263C, 2001A&A...370..157S}. The $L_{\textrm{X}}$/$L_{\textrm{bol}}$ ratio also represents the X-ray activity level of binary systems; this value increases as the orbital period becomes shorter \citep{1996AJ....112.1570P, 2001A&A...370..157S}. Previous studies of the relationship between period and X-ray emission of EWs often suffer from a limited sample size and large scattering of the data, making it difficult to describe them quantitatively, thus impeding further studies of the physical mechanism. Therefore, enlarging the sample size and improving the X-ray coverage and data quality are very important for elucidating the origins and properties of X-ray emission and further investigating the evolution of binary systems. 

\par

In this paper, we identify a large number of EW-type binaries with X-ray emission by cross-matching the All-Sky Automated Survey for Supernovae Variable Stars Database (AVSD) with the 4XMM-Newton Data Release 9 (4XMM-DR9) and the RASS catalogs. Combining with the spectral parameters from the seventh data release of the Large Sky Area Multi object Fiber Spectroscopic Telescope (LAMOST DR7) and binary absolute parameters collected from lectures, we investigate the possible mechanisms of X-ray radiation. In Section~\ref{sec:dataselection}, we describe the selection of our sample. In Section~\ref{sec:data_analysis}, we mainly provide the correlation analyses for X-ray luminosity and activity level versus stellar rotation, along with the spectral and component parameters. We discuss the physical implications of our results in Section~\ref{sec:Discu} and provide a summary of this work in Section~\ref{sec:Summary}.

\par

\section{Sample selection} \label{sec:dataselection}

\subsection{EWXs in 4XMM-DR9 }\label{sec:Sample_selection_XMM}

\par

The All-Sky Automated Survey for Supernovae (ASAS-SN) is a ground-based optical survey regularly scanning the full visible sky with a cadence of between two and three days, with a sensitivity limit down to $V$ $\lesssim$ 17 mag \citep{2018MNRAS.477.3145J}. ASAS-SN discovered new EW binaries by using V-band light curves with a random forest classifier based on 16 Fourier features and 10 other features describing the statistical and mathematical characteristics that the EW binaries are expected to exhibit \citep{2019MNRAS.486.1907J}. Through a cross-matching the variable stars with Gaia DR2 \citep{2018yCat.1345....0G}, 2MASS \citep{2006AJ....131.1163S}, and ALLWISE \citep{2010AJ....140.1868W, 2014yCat.2328....0C}, the AVSD \footnote{https://asas-sn.osu.edu/variables} provides V-bands light curves, the parallaxes, proper motions, photometry, and color or reddening information for most variable sources. Up until September 2021, 76378 objects have been classified as EW-type binaries from the analysis of $\sim$ 660000 variable stars listed in AVSD. We chose these EW-type binaries as our primary catalog (hereafter, ASAS-SN-EW). The 4XMM-DR9 released 550124 unique X-ray sources detected over the 11204 pointed XMM-Newton EPIC observations \citep{2020A&A...641A.136W}. The long period of data accumulation, high sensitivity, and deep exposures make XMM-Newton very suitable for searches of the X-ray counterparts of EW-type binaries, which are often weak X-ray sources. Firstly, we cross-match the ASAS-SN-EW catalog with the 4XMM-DR9 full catalog with a matching radius of $6^{\prime\prime}$. This process leads to 723 unique X-ray sources (1205 observations in total) with XMM-Newton detections, defined as the Parent Group. Secondly, in order to further purify the Parent Group, we cross-matched it with the ATLAS \citep{2018AJ....156..241H} and WISE \citep{2018ApJS..237...28C} catalogs that provide classifications for binary systems. We use a matching radius of $3^{\prime\prime}$, and only the closest object from multiple matches is selected as the counterpart (less than 1.7\%). There are 433 and 261 unique counterparts from these two catalogs, respectively. We further divided these counterparts into Group A and Group B. 
Group A includes 407 objects labeled as close binaries in one or both of the two catalogs.\footnote{CBF or CBH types in ATLAS, EW or EW/EA types in WISE.} The 123 objects in Group B are those labeled as types other than close binaries in both catalogs. The remaining 193 objects from the Parent Group that have counterparts in neither catalogs are named as Group C.

\par
 
\begin{figure}
\centering
\includegraphics[width = 7cm]{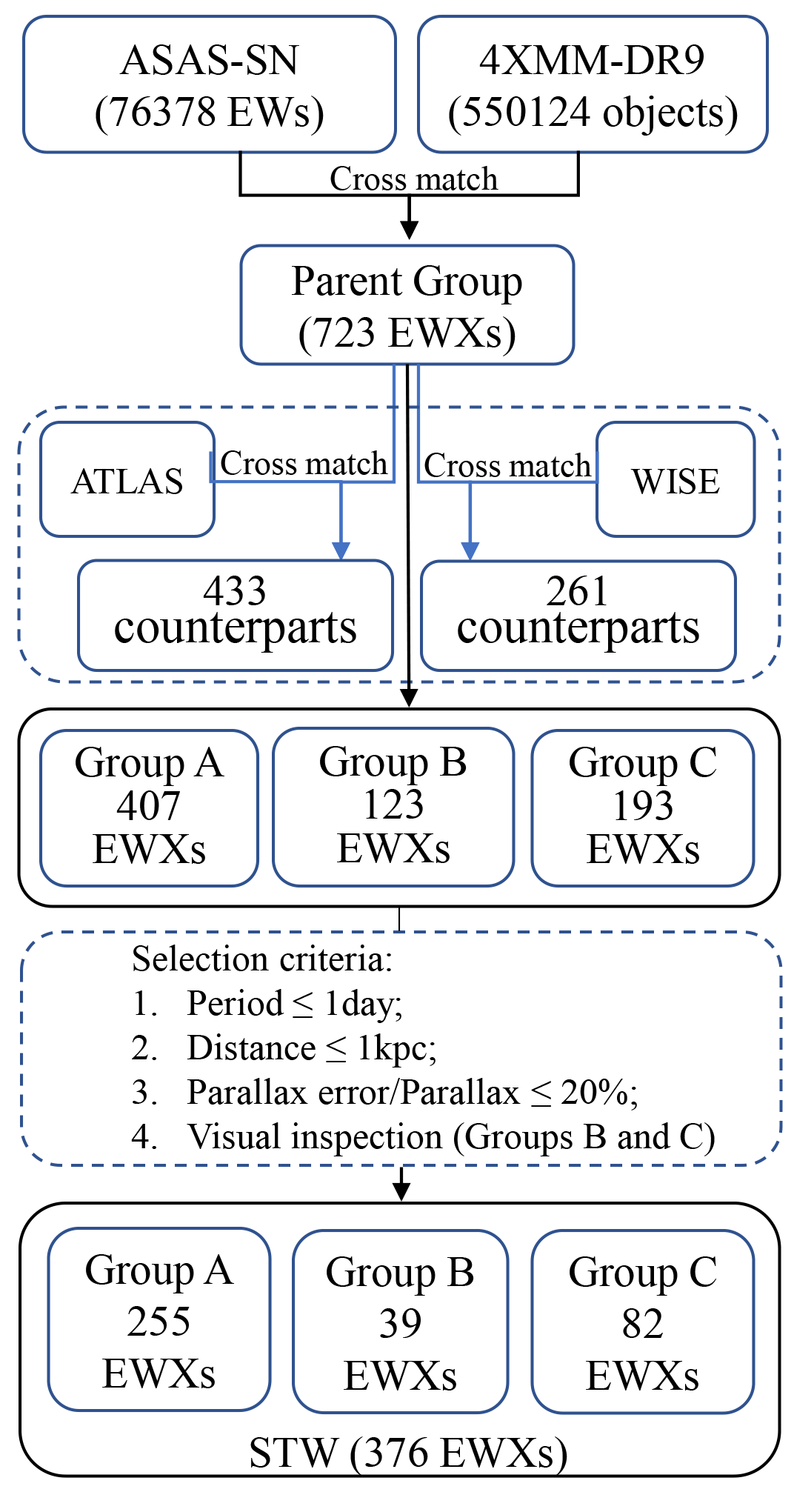}
\caption{Flowchart of compiling STW sample.}
\label{fig:Flowchart}
\end{figure}

We consider the Group A objects as having reliable classifications, since they are consistent in at least two catalogs (one is the ASAS-SN-EW, while the other is either ATLAS or WISE). The objects in Groups B and C were screened again via a visual inspection of their light curves, which is crucial for distinguishing genuine W~UMa binaries from other types of variables that may be misclassified by the mathematical screening criteria used in the production of different catalogs. We calculated the distance of each system in the three groups using Gaia DR2 parallax data and eliminated those with a period $>$ 1~day or distance $>$ 1~kpc (see below), or an uncertainty of distance $>$ 20\% (parallax error/parallax $>$ 20\%). There are 255, 39, and 82 objects retained in Groups A, B, and C, respectively. All these objects combined constitute the main sample of this work (STW hereafter). The full process of this sample compilation is illustrated in the flowchart in Figure~\ref{fig:Flowchart}. The STW contains 376 objects in total, listed in Table~\ref{table:Properties_EWXs}. The columns are organized as ASAS-SN name, right ascension (R.A.; J2000), declination (DEC.; J2000), period, parallax, distance, 4XMM-DR9 name, log$L_{\textrm{X}}$, ATLAS classification, and WISE classification. All the X-ray fluxes of our objects are taken from the 4XMM-DR9, which provides an average unabsorbed flux value in cases of multiple observations for a given source. The X-ray luminosity (log$L_{\textrm{X}}$) is calculated from the full-band X-ray flux in 0.2-12~keV and the Gaia DR2 distance. Almost all of the sources in the STW have $L_{\textrm{X}} > 10^{29}$~erg~s$^{-1}$. 

\par

\begin{sidewaystable*}[!htpb]
  \centering
  \caption{Properties of EWXs in STW (Groups A, B, and C)}
  \label{table:Properties_EWXs}
  \small
  \begin{tabular}{rrrcccrccccrr}
  \hline\noalign{\smallskip}
  ASAS-SN Name & R.A. (J2000) & DEC. (J2000) & Period& Parallax & Distance & 4XMM-DR9 Name & log$L_{\textrm{X}}$ & ATLAS Type & WISE Type & $A_{\rm G}$ & log$L_{\textrm{bol}}$ & log($L_{\textrm{X}}/L_{\textrm{bol}}$) \\
(ASASSN-V) & ($^{\circ}$) & ($^{\circ}$) & (days) &  & (pc) & (4XMM) & (erg s$^{-1}$) &  & & mag & (erg s$^{-1}$) & \\
  \hline\noalign{\smallskip}
                & & & & &  &  Group A & & & & & & \\
        J001322.67+054009.6 & 3.34447 & 5.66933 & 0.28948 & 1.9764 & 498.85 & J001322.7+054008 & 29.738  & EW/EA    & CBF & 0.525 & 33.538 & -3.800\\ 
                J001445.74-391435.4 & 3.69058 & -39.24317 & 0.36436 & 5.3361 & 186.40 & J001445.7-391435 & 30.255  & EW     &   & 0.997 & 33.812 & -3.557\\ 
                ... & ...&...&...&...&...&...&...&...&...&...&...&...\\
                J184128.52+622409.4 & 280.36883  & 62.40270 & 0.28545 & 1.0850 & 898.01 &J184128.5+622408 & 30.222 & EW  &dubious  & 0.600* & 33.655 & -3.433\\
                J195923.73+225703.6   & 299.84887 & 22.95099 & 0.38876 & 1.4855 & 660.76 & J195923.6+225702 & 29.800  & EW               & MSINE& 0.467  & 34.275 & -4.475 \\ 
                J232739.41-004346.6   & 351.91420 & -0.72962 & 0.41252 & 1.8069 & 544.93 & J232739.2-004345 & 30.204  & EW               & SINE  & 0.393 & 34.202 & -3.998\\ 
                 & & & & & & Group B & & & & &    &  \\
                J000525.84-084035.2   & 1.35768 & -8.67644 & 0.26309 & 1.8841 & 523.14 & J000525.8-084035 & 29.879  &      & MPULSE & 0.566 & 33.856 & -3.977\\ 
                J013143.20+302327.9   & 22.93000 & 30.39107 & 0.26660 & 2.2050 & 447.81 & J013143.1+302329 & 29.773  &      & MSINE &  0.702  & 33.464 & -3.691\\ 
                ... & ...&...&...&...&...&...&...&...&...&...&...&... \\
                J232008.39+241055.1   & 350.03494 & 24.18197 & 0.32222 & 1.7312 & 568.21 & J232008.3+241055 & 30.036  &      & MSINE  &0.637  & 33.856 & -3.820\\ 
                J232907.52+145731.2   & 352.28132 & 14.95866 & 0.26678 & 1.8248 & 547.99 & J232907.4+145731 & 29.714  &      & MSINE  &0.720  & 33.516 & -3.802\\
                & & & & & & Group C & & & &  &    & \\
                J002150.83-704642.5   & 5.46179 & -70.77846 & 0.27176 & 1.9298 & 510.58 & J002150.5-704640 & 29.963 &  &  &0.639  & 34.025 & -4.062\\ 
                J002234.47+614417.2   & 5.64364 & 61.73811 & 0.35770 & 1.3647 & 717.89 &J002234.3+614417 & 30.032 &  &  &1.208 &  34.107 & -4.075\\ 
                ... & ...&...&...&...&...&...&...&...&...&...&...&...\\
                J234658.78-545310.2   & 356.74490 & -54.88617 & 0.28150 & 2.6239 & 376.96 & J234658.8-545308 & 29.719 &  &   &0.531  & 33.475 & -3.756\\ 
                J235142.03-395949.8   & 357.92512 & -39.99716 & 0.28395 & 1.2965 & 754.65 & J235142.1-395946 & 29.892 &  &   &0.475  & 33.777 & -3.885\\ 

  \hline\noalign{\smallskip}
  \end{tabular}  
  \flushleft
      \begin{tablenotes}
     \footnotesize
    \item[1] (This table is available in its entirety in the online machine-readable form.)
    \end{tablenotes}
 \end{sidewaystable*}

\par

\par

One factor which could potentially affect the accuracy of X-ray luminosity is the presence of X-ray flares. Since there has been sparse mention in the literature on this aspect of EWXs, we refer to the occurrence frequency of X-ray flares for main sequence stars. \citet{2019A&A...628A..41P} selected 102 X-ray emitting main sequence stars with spectral types ranging from A to M to investigate their magnetic activities and X-ray flares. They detected six flares on five stars, indicating a low X-ray flare rate. Therefore, we did not consider the effects of X-ray flares in our analysis.

\par

\subsection{EWXs in RASS} \label{sec:sec:Sample_selection_RASS}

\par

By reprocessing the RASS data, \citet{2016A&A...588A.103B} released an updated catalog with 135118 X-ray sources (the 2RXS catalog). We compiled a sample of 190 EWXs detected by RASS following the same procedure as that in Section~\ref{sec:Sample_selection_XMM}, but with the X-ray-to-optical matching radius set to $20^{\prime\prime}$. 
We follow \citet{2001A&A...370..157S} and \citet{2006AJ....131..990C} to convert the count rate to X-ray flux using the energy conversion factor.  The X-ray luminosity is calculated with the Gaia DR2 distance.

\par

In addition, we collected the EWXs identified with RASS from the literature and updated their distances and X-ray luminosities using the Gaia DR2 distances. \citet{2001A&A...370..157S} found that log($L_{\textrm{X}}$/$L_{\textrm{bol}}$) declines with the decreasing color index $B-V < 0.6$. \citet{2006AJ....131..990C} investigated the relationship between the X-ray luminosity and the stellar parameters, such as the rotation period and color index. In combining their sample with those of \citet{2001A&A...370..157S} and \citet{1996MNRAS.280..627M}, these authors reinforced a relation in which the shorter the rotation period, the weaker the X-ray emission when the rotational period is shorter than $\sim0.5$ days. \citet{2006AJ....131..633G} also used the RASS database to estimate the X-ray emissivity of contact binaries in our Galaxy. These previous studies utilized a variety of approaches to calculate the distance of each binary. \citet{2006AJ....131..633G} used a period-color-luminosity relation with $J-H$ colors. The distances of the \citet{2006AJ....131..990C} sample were obtained via the absolute magnitude calibrated with period and $B-V$ colors, while those of \citet{2001A&A...370..157S} were derived from the parallaxes released by the Hipparcos satellite. In total, they identified 231 EWXs (including duplicates; for more details,  see below). 
We cross-matched these 231 objects with Gaia DR2 using the $3^{\prime\prime}$ radius to obtain the updated parallaxes. For cases with multiple matches (about 18 \%), the closest object is selected as the counterpart. In order to ensure the consistency across different samples, we only kept objects with parallax values in Gaia DR2. In total, there are 54/57 objects from \citet{2001A&A...370..157S}, 117/140 from \citet{2006AJ....131..633G}, and 9/34 from \citet{2006AJ....131..990C} remaineing. The X-ray luminosity of each object in these three groups was updated with the new distance from Gaia parallax. Duplicated objects among these three samples were removed. We finally obtained 165 EWXs from the literature (54 from \citealt{2001A&A...370..157S}, 106 from \citealt{2006AJ....131..633G} and 8 from \citealt{2006AJ....131..990C}).

\par

Finally, after combining the EWXs we obtained from 2RXS catalog and those from the literature and then removing 36 duplicated objects, we obtained a sample of 319 objects consistently selected from RASS, named as the SRASS sample (listed in Table~\ref{table:SRASS_EWXs}). All the EWXs in SRASS are within 1~kpc from us. The STW objects are also limited within 1~kpc for consistency. Because of the substantially different X-ray detection sensitivity between 4XMM and RASS, we could not merge these two samples. Instead, we performed the same analyses for the two samples separately and compared their results, as we present in the following sections. Eight sources are duplicated in STW and SRASS, and each of them has similar X-ray luminosities in both catalogs. In order to maintain the completeness of the samples, we kept them in their respective datasets in the analysis.

\par
 \begin{table*}[!htpb]
   \caption{Properties of objects in SRASS}
     \label{table:SRASS_EWXs}
   \begin{center}
   \begin{tabular}{rrccclccc}\hline \hline

R.A. (J2000)  &DEC. (J2000) & Period & Distance & log$L_{\textrm{X}}$ & $A_{\rm G}$ & log$L_{\textrm{bol}}$  & log($L_{\textrm{X}}$/$L_{\textrm{bol}}$) & Reference   \\
($^{\circ}$) & ($^{\circ}$)       & (days) & (pc)         & (erg s$^{-1}$)            & mag          & (erg s$^{-1}$)       & &    \\ \hline
324.70584 &  26.69278 & 0.28040 & 152.54 & 29.880 & 0.612 & 33.689 & -3.809 & (1)   \\
150.41876 &  17.40888 & 0.28410  & 62.87 & 30.188  & 1.211  & 33.883 & -3.695 & (1)   \\
...&...&...&...&...&...&...&...&... \\
330.69898 & -12.31137 & 0.30678 & 182.27 & 30.261 & 0.454 & 33.732 & -3.471 & (5)    \\
347.74778 & 21.71198 & 0.25817 & 105.88 & 30.168 & 1.058 & 33.505 & -3.337 & (5)  \\
\hline\noalign{\smallskip}
  \end{tabular}
  \end{center}
  \begin{tablenotes}
     \footnotesize
    \item[1] Ref.(1): \citet{2001A&A...370..157S}, (2):\citet{2006AJ....131..633G}, (3):\citet{2006AJ....131..990C}, (4): Duplications in STW, (5): This Work.
    \item[2] (This table is available in its entirety in the online machine-readable form.)
    \end{tablenotes}
\end{table*}

\par

\subsection{Spectra from LAMOST}\label{sec:Spectra_of_LAMOST}

\par

We utilized the LAMOST stellar spectral database to investigate the relationships between the optical spectral parameters and X-ray properties of EW-type binaries. 
The LAMOST is a spectroscopic survey telescope located at the Xinglong station, National Astronomical Observation of China, with a $\sim4$-meter effective aperture and a field of view of $5^\circ$ \citep{2012RAA....12.1197C}. LAMOST can obtain 4,000 spectra covering wavelength from 3700 to 9000 {\AA} with a resolving power of 1800 in a single exposure. We input the coordinates of STW and SRASS samples into the LAMOST DR7 database\footnote{http://dr7.lamost.org} with a matching radius of $3^{\prime\prime}$ to obtain stellar spectral parameters, including the effect temperature, $T_{\rm eff}$, surface gravity, $\log g$, metallicity [Fe/H], and radial velocity, $V_{\rm r}$; these values were automatically derived by the LAMOST stellar parameters pipeline \citep{2011RAA....11..924W, 2014IAUS..306..340W}.  A total of 139 unique sources are matched in LAMOST DR7 (hereafter, the Spec-EWX sample), which are listed in Table~\ref{table:Spectra_EWXs}. The designation of each column is defined as follows: right ascension (J2000), declination (J2000), log$L_{\textrm{X}}$, period, effective temperature ($T_{\rm eff}$), error of $T_{\rm eff}$, metallicity ([Fe/H]), error of [Fe/H], surface gravity ($\log g$), and its error.

\par
\begin{table*}[!htpb]
  \centering
  \caption{Stellar spectral parameters of EWXs in the Spec-EWX sample}
  \label{table:Spectra_EWXs}
  \small
  \begin{tabular}{rrcccccccc}
  \hline\noalign{\smallskip}
R.A. (J2000) & DEC. (J2000) & log$L_{\textrm{X}}$ & Period  &$T_{\rm eff}$ & $T_{\rm eff}$ error & [Fe/H] &  [Fe/H] error & $\log g$  &$\log g$ error \\
($^{\circ}$) & ($^{\circ}$) & (erg s$^{-1}$) & (days) & (K) & (K) & (dex) &  (dex) & (dex) & (dex) \\
  \hline\noalign{\smallskip}
283.39054 &47.43649 &30.213 &0.31503 &5589.50&21.82&-0.10&0.02&4.29& 0.04 \\ 
8.52526 &39.69754 &29.078 &0.28669 &5220.21&63.34&0.25&0.06&4.43& 0.10 \\ 
...&...&...&...&...&...&...\\
349.62904 &42.92814 &30.072 &0.34180 &5818.81&25.68&0.09&0.02&4.20& 0.04 \\ 
350.03494 &24.18197 &30.036 &0.32222 &5416.84&32.06&0.19&0.03&4.291& 0.05 \\
  \hline\noalign{\smallskip}
  \end{tabular}  
  \flushleft
    \begin{tablenotes}
     \footnotesize
    \item[1] (This table is available in its entirety in the online machine-readable form.)
    \end{tablenotes}
\end{table*}

\par

\subsection{Upper limits for X-ray nondetections} \label{sec:sec:upper limits}

For the completeness of the sample selection, we determined that EW binaries with X-ray nondetections should also be considered. We screened the ASAS-SN-EW samples following the same procedures as in Section~\ref{sec:Sample_selection_XMM} (corresponding to the ``Group A" objects after applying the cuts on period, distance, and parallax precision) without matching X-ray source catalogs. A total of 14096 reliably-classified EW binaries are obtained. After removing the X-ray detected objects we already obtained in Section~\ref{sec:Sample_selection_XMM}, we calculated the X-ray flux upper limits using the web client {\it HIgh-energy LIght curve GeneraTor} (\citealt{2022A&C....3800531S})\footnote{http://xmmuls.esac.esa.int/hiligt/scripts/
hiligt.py} which can poll individual servers for the chosen X-ray missions, and return the X-ray flux and/or upper limits for given targets or coordinates.
The default parameter settings were adopted, namely: a 2$\sigma$ upper limit significance, an absorbed power law spectral model for flux conversion with a photon index  of $\Gamma=2,$ and the hydrogen column density of $N_{\rm H} =3 \times 10^{20}\ \rm cm^{-2}$. The XMM-Newton observations in both slew mode \citep{2012A&A...548A..99W} and pointed mode were searched, while the entire RASS was utilized. When multiple upper limits were returned for the same coordinate, we chose the lower value (to achieve a tighter constraint). In particular, we used the upper limits from pointed observations for XMM-Newton rather than those from the slew mode when both exist. Finally, we get 12279 (39 from pointed mode and 12240 from slew mode) and 13769 X-ray flux upper limits from the XMM-Newton and ROSAT servers, respectively. Combining the distance data of Gaia DR2, we further calculated the X-ray luminosity upper limits, listed in Tables~\ref{table:XMMnewtonupperlimits} and \ref{table:ROSATupperlimits} for XMM-Newton and ROSAT, respectively. These X-ray upper limits are also considered in the subsequent analyses.

\par

 \begin{table*}[!htpb]
   \caption{Properties of EW-type binaries with X-ray luminosity upper limits from XMM-Newton pointed and slew surveys}
   \begin{center}
   \begin{tabular}{rrcccc}\hline \hline

R.A. (J2000) & DEC. (J2000) & Modes & Period & Distance & log$L_{\textrm{X}}$\\
($^{\circ}$)   & ($^{\circ}$)     &             & (days)    & (pc)       & (erg s$^{-1}$)  \\ \hline
  0.01355 & 69.37062 & slew & 0.37280 & 902.25 & $<$32.072 \\
  0.14513 & 53.81984 & slew  & 0.23166  & 583.38 & $<$31.536 \\
... & ... & ... &... &... & ...  \\
359.97490 & 66.15284 & slew  & 0.32750 & 835.48 & $<$31.985 \\
359.98450 & 51.50363  & slew & 0.29762 & 579.29 & $<$31.561 \\

\hline\noalign{\smallskip}
  \end{tabular}
  \end{center}
  \label{table:XMMnewtonupperlimits}
    \begin{tablenotes}
     \footnotesize
    \item[1] (This table is available in its entirety in the online machine-readable form.)
    \end{tablenotes}
\end{table*}

\par
 \begin{table*}[!htpb]
   \caption{Properties of EW-type binaries with X-ray luminosity upper limits from the ROSAT survey}
   \begin{center}
   \begin{tabular}{rrcccc}\hline \hline

R.A. (J2000) & DEC. (J2000) & Modes & Period & Distance & log$L_{\textrm{X}}$\\
($^{\circ}$)   & ($^{\circ}$)     &             & (days)    & (pc)       & (erg s$^{-1}$)  \\ \hline
   0.01355 & 69.37062& RosatSurvey& 0.37280 &902.25 & $<$31.385 \\
    0.14513  &53.81984 &RosatSurvey &0.23166 &583.38 & $<$30.956 \\
... & ... & ... &... &... & ...  \\
359.97490 & 66.15284& RosatSurvey &0.32750 &835.48 & $<$31.492 \\
359.98450  &51.50363 &RosatSurvey &0.29762 &579.29 & $<$31.047 \\

\hline\noalign{\smallskip}
  \end{tabular}
  \end{center}
  \label{table:ROSATupperlimits}
    \begin{tablenotes}
     \footnotesize
    \item[1] (This table is available in its entirety in the online machine-readable form.)
    \end{tablenotes}
\end{table*}

\par

\section{Data analysis} \label{sec:data_analysis}

\subsection{X-ray emission versus the period} \label{sec:X-ray_P}

\par

\begin{figure*}
\centering
\includegraphics[width = 15cm]{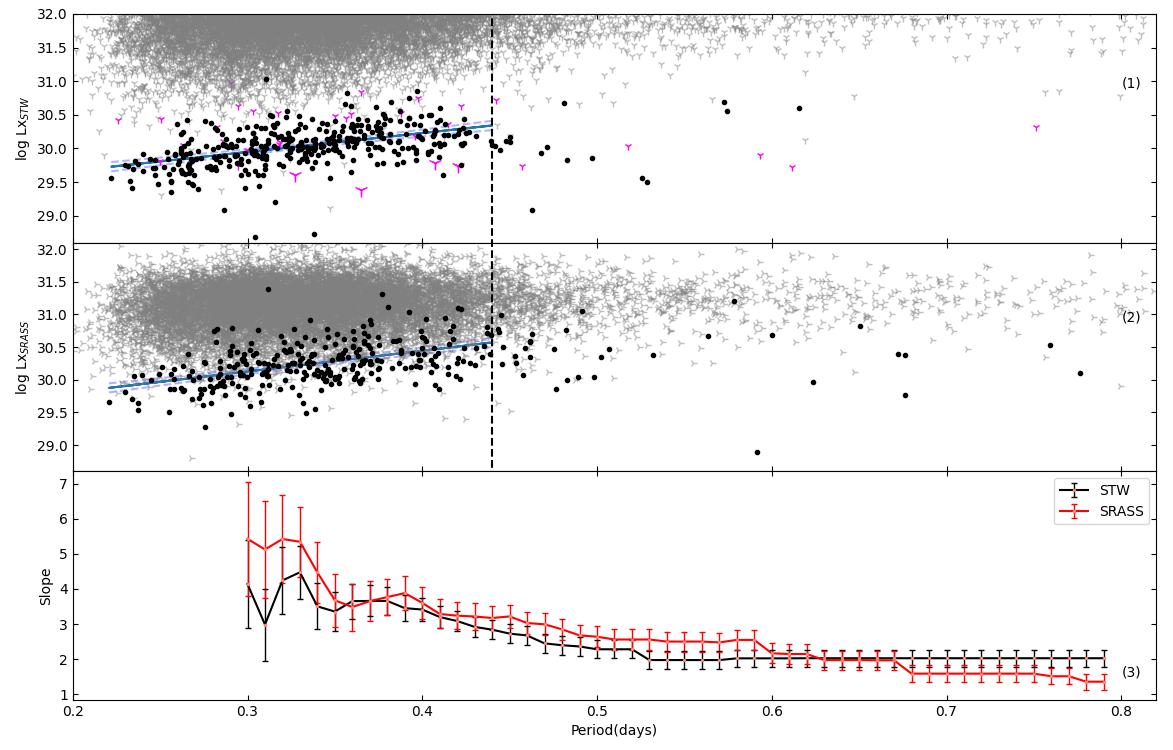}
\caption{Orbital period vs. log$L_{\textrm{X}}$: for STW (1) shown as black points. The magenta and gray `$tri\_down$' symbols represent the X-ray upper limits for some EW-type binaries from XMM-Newton pointed and slew survey, respectively;  for SRASS (2)  shown in black points. The gray `$tri\_right$' symbols represent the X-ray upper limits for some EW-type binaries from ROSAT survey. The period vs. slopes with their corresponding error bars of STW (black) and SRASS (red) with different limiting period ranges (3). The black dashed lines are at $P=0.44$ days. The solid blue lines stand for best-fit linear correlations, and the light blue dashed lines represent 95\% uncertainty ranges.}
\label{fig:Lx_P}
\end{figure*}

In this section, we investigate the relationship between the X-ray emission and rotation for EW-type binaries. Contact binaries have circular orbits and synchronous co-rotation, which mean that the orbital period of the binary equals to the spin period of each component. In Figure~\ref{fig:Lx_P} (1) and (2), we plot the period against log$L_{\textrm{X}}$ of the EWXs in STW and SRASS, respectively. Objects with X-ray detections and upper limits are both plotted. Most of the upper limits do not provide physically meaningful constraints because of the low sensitivity. We discuss those X-ray nondetections in Section~\ref{sec:completeness}. For the X-ray detected EW binaries, the X-ray luminosity has a significant positive correlation with the orbital period of the binary stars between $\sim$0.2 and $\sim$0.45 days. Meanwhile, in the period interval [0.4,0.5], there appears to be a break point (designated ``P1") after which the correlations between the period and log$L_{\textrm{X}}$ no longer hold, and the data points become more scarce and scattered. 

\par

\par

As shown in panel (3) of Figure~\ref{fig:Lx_P}, the value of each point on the black and red lines represents the slope $a$ of the least-squares quadratic fitting (in the form of $\log L_{\rm X} = a \times P + b $) to all objects with a period less than its abscissa value  \textit{P}. The black broken line with error bars is for STW, while the red one line is for SRASS. The fitting range is from 0.3 to 0.8 days with the step of 0.01 days. We performed the Kendall$^{\prime}$s $\tau$ test \citep{1990BJpc....25..86} between \textit{P} and log$L_{\textrm{X}}$ for objects in these two samples. The null-hypothesis (i.e., no correlation exists) probability is always less than 10$^{-5}$ for the period range of [0.36, 0.80], which suggests that the periods of EWXs show a strong linear correlation ($>5\sigma$) with the log$L_{\textrm{X}}$ for both samples.
With the increasing period (and thus the number of objects), the slopes of the two samples gradually converge. The fitting slopes of STW and SRASS are both within the 1$\sigma$ uncertainty range of each other, showing that these two samples are highly consistent with each other in the relationship between \textit{P} and $\log L_{\rm X}$. Especially in the period range of [0.38, 0.44], the difference between the slopes is less than 1$\sigma$. Beyond $P=0.44$ days, the slope difference starts to increase as the period gets longer, which means that P1 is likely to be at 0.44 days. Meanwhile, only a few data points have a period greater than 0.44 days. Therefore, we took 0.44~days as the upper bound for the period of STW ($\sim$96\% of its objects) and SRASS ($\sim$90\%) samples for the correlation analysis; that is, data points with period values less than 0.44 days were selected for the least-squares fitting. The blue lines in panels (1) and (2) of Figure~\ref{fig:Lx_P} represent the best-fit relation and the light blue dashed lines represent the 95\% uncertainty range. We formulate, for the first time, a clear linear correlation between the period $P$ (log$P$) and X-ray luminosity log$L_{\textrm{X}}$ for EWXs, which is described by the following Equations~(\ref{equ:STW_LX_P})--(\ref{equ:SRASS_LX_logP}) for the STW and SRASS: 

\par

\begin{equation}\label{equ:STW_LX_P}
    \log L_{X_{STW}} = 2.81(27) \times P + 29.10(9),
\end{equation}

\begin{equation}\label{equ:SRASS_LX_P}
    \log L_{X_{SRASS}} = 3.17(35) \times P + 29.17(12),
\end{equation}

\begin{equation}\label{equ:STW_LX_logP}
    \log L_{X_{STW}} = 2.14(24) \times \log P + 31.08(10),
\end{equation}

\begin{equation}\label{equ:SRASS_LX_logP}
   \log L_{X_{SRASS}} = 2.45(27) \times \log P + 31.41(13).
\end{equation}

\par 

While the dependency of X-ray luminosity upon period has been pointed out by previous studies (e.g., \citealt{2001A&A...370..157S}, \citealt{2006AJ....131..990C}, and \citealt{2006AJ....131..633G}), quantitative descriptions had not been provided. The linear correlations between orbital period and X-ray luminosity for the STW and SRASS samples we present here are consistent with each other ($\simlt1\sigma$). The use of Gaia's more accurate parallax information, which results in more accurate log$L_{\textrm{X}}$ values in our work help reduce the scatter of the data, making the correlations more clear and better defined. The linear correlations shown in both the STW and SRASS have high statistical significance: the probabilities of null-hypothesis (i.e., no correlation exists) in both cases are $<10^{-5}$, corresponding to a $>5\sigma$ confidence level. For EWs (regardless of their X-ray emission), \citet{2018ApJ...859..140C} has established the positive correlation between period and luminosity in the optical and mid-infrared bands. The period of their sample ranges from 0.25 to 0.56 days. As a comparison, the positive correlation between period and X-ray luminosity is well maintained at $P=0.2$ to 0.44 days. 

\par

\begin{figure*}
\centering
\includegraphics[width = 15cm]{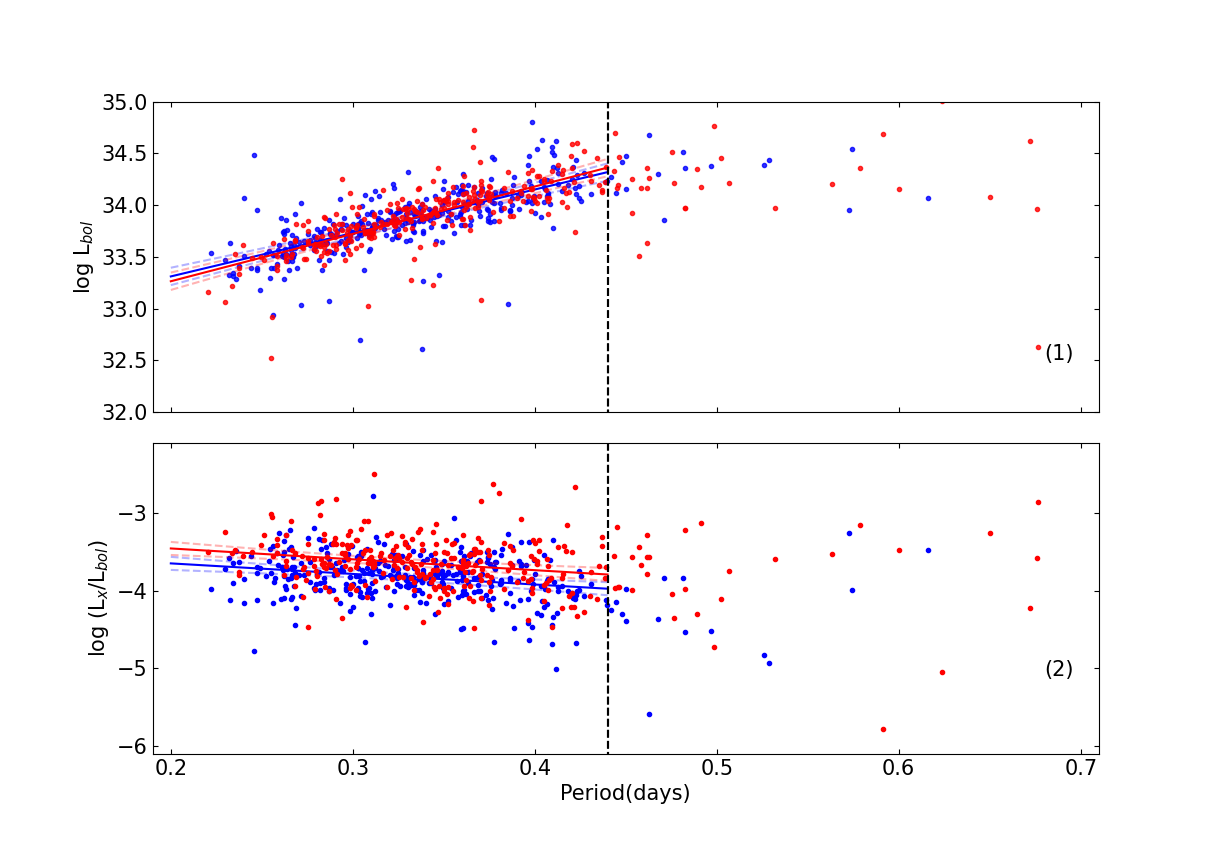}
\caption{Orbital period vs. log$L_{\textrm{bol}}$: for STW in blue points and SRASS in red points (1).  Orbital period vs. log$(L_{\rm X}/L_{\rm bol})$ for STW in blue points and SRASS in red points (2). Black dashed lines are at $P=0.44$ days. Solid blue and red lines stand for best-fit linear correlations ($P<0.44$~days) for STW and SRASS respectively, while the light blue and red dashed lines are their corresponding 95\% uncertainty ranges.}
\label{fig:Lx_Lbol_STW_sub_SRASS}
\end{figure*}

Based on the method provided by \citet{2018A&A...616A...8A}, we can calculate the bolometric luminosity $L_{\textrm{bol}}$ of the EWXs with the following equations:
\begin{equation}\label{equ:bolometricL}
    -2.5~\textrm{log}_{10}~(L_{\textrm{bol}}/L_{\odot}) = M_{\textrm{G}} + BC - M_{\rm sun},
\end{equation}
 \begin{equation}\label{equ:Mg}
    M_{\textrm{G}} = G + 5 - 5\log_{10} D - A_{\rm G},
\end{equation}
where $M_{\textrm{G}}$ is the absolute $G$-band magnitude, $BC$ is the temperature-dependent bolometric correction (based on the effective temperatures in Gaia DR2 estimated from the $BP$-band, $RP$-band, and $G$-band magnitudes; see \citealt{2018A&A...616A...8A} for details), $M_{\rm sun}$ is the solar bolometric magnitude 4.74~mag, $G$ is the apparent magnitude in $G$-band, $D$ is the distance, and $A_{\rm G}$ is the $StarHorse$ extinction \citep{2019A&A...628A..94A} in the $G$-band.\footnote{For sources without this value, we set it to 0.600 mag according to the statistical distributions of extinction and distance of our samples.} The values of $A_{\rm G}$, log$L_{\rm bol}$, and log$(L_{\rm X}/L_{\rm bol})$ of 370 EWXs from STW are listed in the last three columns of Table~\ref{table:Properties_EWXs}, while those of 318 EWXs from SRASS are listed in the Columns (6)--(8) of Table~\ref{table:SRASS_EWXs}; for the remaining 6 sources in STW and 1 source in SRASS, these three parameters are not calculated because of the lack of $BP$/$RP$-band coverage. In Figure~\ref{fig:Lx_Lbol_STW_sub_SRASS} (1) and (2), we plot the period against log$L_{\textrm{bol}}$ and log$(L_{\rm X}/L_{\rm bol})$ of the EWXs in STW and SRASS with blue and red symbols, respectively. The blue and red solid lines represent the best-fit linear relations ($P<0.44$~days), while the light blue and light red dashed lines represent the 95\% uncertainty ranges for these two samples, respectively.  The Kendall$^{\prime}$s $\tau$ tests suggest that the confidence levels of $P$ correlated with both the log$L_{\textrm{X}}$ and log$(L_{\rm X}/L_{\rm bol})$ are all greater than 5~$\sigma$. The fitting parameters are listed in Table~\ref{table:ste_sub_srassLX/Lbol}.

\par

\par

\begin{table}[!htpb]
   \caption{Best-fit parameters for $P$-log$L_{\textrm{bol}}$ and $P$-\-log($L_{\textrm{X}}$/$L_{\textrm{bol}}$) of STW and SRASS}
   \begin{center}
   \begin{tabular}{llrrrr}\hline \hline
Relations &  Samples & Slope & Intercept  \\
\hline \\
  $P$-log$L_{\textrm{bol}}$               & STW              &    4.19 $\pm$ 0.21 &  32.47 $\pm$ 0.07  \\
                          & SRASS &  4.57 $\pm$ 0.23 &  32.34 $\pm$ 0.08  \\ 
$P$-log($L_{\textrm{X}}$/$L_{\textrm{bol}}$)               &   STW     &  -1.36 $\pm$ 0.29 &  -3.37 $\pm$ 0.10  \\
                          & SRASS&   -1.39 $\pm$ 0.37 &  -3.17 $\pm$ 0.13  \\      
\hline\noalign{\smallskip}
  \end{tabular}
  \end{center}
  \label{table:ste_sub_srassLX/Lbol}
\end{table}

\par

\subsection{Spectral parameters analysis}\label{sec:Spectral_parameters}

\par

\begin{figure*}
\centering
\includegraphics[width = 15cm]{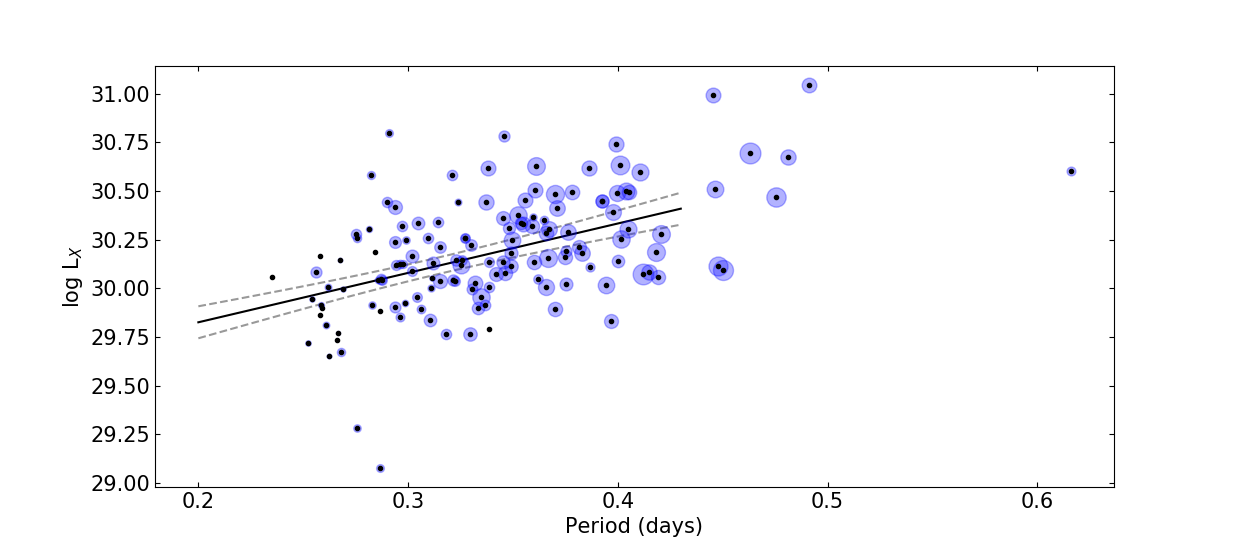}
\caption{Orbital period vs. log$L_{\textrm{X}}$ of Spec-EWX sample (black points) with LAMOST stellar parameters. Solid black line shows the least-squares linear fitting for the data with a period less than 0.44 days. Light black dashed lines represents 95\% uncertainty range. The size of the light blue circles represents the binary temperature ($\sim4500$--6600~K). }
\label{fig:LAMOST_P_Lx}
\end{figure*}

The Spec-EWX sample has stellar spectral parameters measured from LAMOST spectra. First, we sought to address whether they indeed constitute a representative sample of the STW and SRASS.  
We performed K-S tests on the distributions of period between Spec-EWX and these two samples; the probability $P_{\rm K-S}$ are 93.2\% and 22.1\%, respectively, which suggest they follow the same distributions. As shown in Figure \ref{fig:LAMOST_P_Lx}, we also provide the linear fitting for the period versus log$L_{\textrm{X}}$ relation (black points) for Spec-EWX sample with $P<0.44$~days. The best-fit slope and intercept values are 2.53(46) and 29.31(16), respectively, which is consistent within $\sim 1.5\sigma$ error of those values in Equations (\ref{equ:STW_LX_P}) and (\ref{equ:SRASS_LX_P}).

\par

\par

\begin{figure*}
\centering
\includegraphics[width = 18cm]{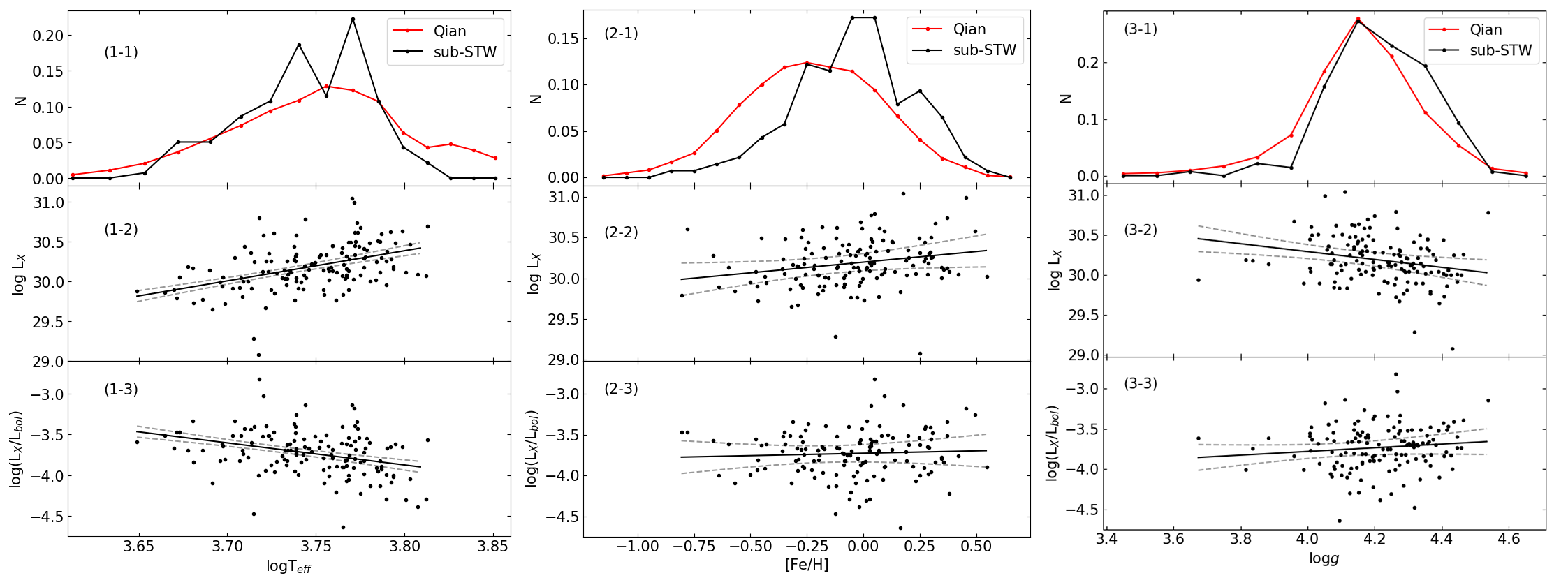}
\caption{Normalized distributions of three spectral parameters ($\log T_{\rm eff}$, [Fe/H] , and $\log g$) of EWXs, and their relationships with log$L_{\textrm{X}}$ and log($L_{\textrm{X}}$/$L_{\textrm{bol}}$). Upper panels: Normalized distributions of the binary temperature (1-1), metallicity [Fe/H] (2-1), and surface gravity $\log g$ (3-1). The red and black broken lines represent the EW-type binaries collected by \citet{2017RAA....17...87Q} and our Spec-EWX sample, respectively. 
Middle panels: Relationships between  log$L_{\textrm{X}}$ and the binary temperature (1-2), metallicity [Fe/H] (2-2), and surface gravity $\log g$ (3-2). 
Lower panels: Relationships between log($L_{\textrm{X}}$/$L_{\textrm{bol}}$) and the binary temperature (1-3), metallicity [Fe/H] (2-3), and surface gravity $\log g$ (3-3).
Solid black lines are the least-squares linear fits and the light black dashed lines represent the 95\% uncertainty ranges.}
\label{fig:LAMOST_logT_logg_FeH_Lx_N}
\end{figure*}

The statistical distributions of binary temperature $\log T_{\rm eff}$, metallicity  ([Fe/H]), and the surface gravity $\log g$ are shown in the upper panels of Figure~\ref{fig:LAMOST_logT_logg_FeH_Lx_N} (1-1, 2-1, 3-1). For all of the Spec-EWX sample, the effective temperature, $T_{\rm eff}$, ranges from $\sim4450$~K (K5) to $\sim6600$~K (F2), while the majority are distributed between 5150~K and 6000~K (corresponding to $\sim$ K0V to G0V spectral type), indicating that they are solar-like main sequence stars. The metallicity [Fe/H] ranges from $\sim-1.00$ to $\sim0.6$ dex, with a mean value of -0.05 dex. The distribution peaks at $-0.05$ dex which is slightly lower than that of Sun, that is, [Fe/H] = 0. The surface gravity $\log g$ ranges from $\sim3.4$ to $\sim4.7$, with the mean value of 4.22 and the distribution peak at $4.15$.  We also plot the normalized distributions of $\log T_{\rm eff}$, [Fe/H], and $\log g$ of EW-type binaries collected by \citet{2017RAA....17...87Q} in panel (1-1), (2-1) and (3-1) of Figure~\ref{fig:LAMOST_logT_logg_FeH_Lx_N}, respectively. K-S tests between our Spec-EWX  and their sample were performed for these three parameters. The values of $P_{\textrm{K-S}}$ are 2.72 $\times 10^{-4}$, 7.25$\times 10^{-10}$, and 4.63$\times 10^{-5}$, respectively, suggesting that for each of the three parameters, the statistical distributions of these two samples differ at the $> 3\sigma$ level. 

\par

\begin{table*}[!htpb]
   \caption{Best-fit parameters for log$T$, [Fe/H] and $\log g$ with log$L_{\textrm{X}}$ and log($L_{\textrm{X}}$/$L_{\textrm{bol}}$) of Figure \ref{fig:LAMOST_logT_logg_FeH_Lx_N}, and corresponding correlation probabilities from the Kendall$^{\prime}$s $\tau$ tests}
   \begin{center}
   \begin{tabular}{llrrrrr}\hline \hline
Relations &Parameters&  Slope & Intercept &  $\tau$ & $1-P_{\tau} $ \\
\hline \\
log$L_{\textrm{X}}$& log$T$ &     3.78 $\pm$ 0.65 &  16.01 $\pm$ 2.44  &   0.320  &   $> 99.99\%$  \\
                            & $[\textrm{Fe}/\textrm{H}]$&    0.26 $\pm$ 0.09 &  30.19 $\pm$ 0.02  &      0.188   & 99.89\%    \\ 
                            & $\log g$&    $-$0.49 $\pm$ 0.17 &  32.26 $\pm$ 0.72     &    $-$0.198 & 99.94\%      \\
                            
log~($L_{\textrm{X}}$/$L_{\textrm{bol}}$)& log$T$ &     $-$2.70 $\pm$ 0.67 &  6.41 $\pm$ 2.50  &  $-$0.232  &   $> 99.99\%$  \\
                            & $[\textrm{Fe}/\textrm{H}]$&    0.06 $\pm$ 0.09 &  $-$3.73 $\pm$ 0.02  &      0.047  & 59.09\%    \\ 
                            & $\log g$&    0.23 $\pm$ 0.17 &  $-$4.69 $\pm$ 0.72  &     0.083 & 84.79\%      \\                           
\hline\noalign{\smallskip}
  \end{tabular}
  \end{center}
  \label{table:leastsquare_parameters}
\end{table*}

\par

\par

The component stars of an EW binary, especially the more massive primary, still maintain the mass-temperature relation of single main sequence stars \citep{2005ApJ...629.1055Y, 2018MNRAS.473.5043E}. Therefore, it is reasonable to use spectral temperature to infer the mass of the primary star of an EW. In terms of temperature, as shown in Figure \ref{fig:LAMOST_logT_logg_FeH_Lx_N} (1-1), both the EWXs and EWs are mainly distributed ($\sim$67\% vs. 48\%) between $\sim5150$~K (corresponding to $\sim$0.79 M$_{\odot}$ for a main sequence star) and $\sim6000$~K ($\sim$1.1 M$_{\odot}$). In particular, the ratio of EWXs with spectral temperatures below 6000~K is 82.7\%, which may indicate that a large fraction of EWXs have primary stars with masses lower than 1.1~M$_{\odot}$. However, EWXs have a higher percentage of objects in this range ($T<6000$~K) than the full EW-binary population has, while the opposite is true for the case of $T>6000$~K. Meanwhile, only a few EWXs have temperatures higher than 6500~K, which indicates that few EWXs have primary stars heavier than $\sim1.4$ M$_{\odot}$.

\par

In Figure~\ref{fig:LAMOST_P_Lx}, the size of the light blue circles around each data point represents the effective temperature of Spec-EWX sample from LAMOST spectroscopy. This plot suggests the positive correlations between X-ray luminosity and both period and effective temperature. However, some sources with a period from $\sim0.40$ to 0.44 days have high temperature but low X-ray luminosity. This can be explained by the general trend of X-ray luminosity exhibited by single main sequence stars. With the ascending temperatures from K to G type stars, the typical X-ray luminosity increases gradually from $\sim$10$^{29}$ erg s$^{-1}$ to $\sim$10$^{30}$ erg s$^{-1}$. However, this trend does not continue to F-type stars as their typical X-ray luminosity ($\sim$10$^{30}$ erg s$^{-1}$) is comparable to that of G-type stars \citep[see][the left panel of Figure 6]{2020ApJ...902..114W}.

\par

As shown in Figure \ref{fig:LAMOST_logT_logg_FeH_Lx_N} (2-1), the peak value of Spec-EWX metallicity distribution is at $\sim-0.05$. About 77\% of the Spec-EWX objects have [Fe/H] $> -0.25$. In contrast, for general EWs (represented by the red line), the distribution peak is at  $\sim-0.25$; the population is distributed almost evenly around this peak value (53\% vs. 47\%). This indicates that EWs with higher metallicity are more likely to produce X-ray emission. 

\par

In Figure~\ref{fig:LAMOST_logT_logg_FeH_Lx_N} (3-1), we can see that the surface gravity of EWXs and of EWs shows nearly the same distribution peaks ($\sim$4.15). At the interval of $4.2\leq \log g \leq 4.5$, the percentage of EWXs is slightly higher than that of EWs. 

\par

The middle and bottom rows of Figure~\ref{fig:LAMOST_logT_logg_FeH_Lx_N} demonstrates the correlations between the above three stellar parameters and X-ray luminosity log$L_{\textrm{X}}$, and X-ray activity level log($L_{\textrm{X}}$/$L_{\textrm{bol}}$) of EWXs, respectively. Linear regressions and Kendall$^{\prime}$s $\tau$ tests were performed and the results are listed in Table~\ref{table:leastsquare_parameters}. The binary temperature $\log T_{\rm eff}$ has a strong positive correlation with log$L_{\textrm{X}}$ at $>5\sigma$. The metallicity is also positively correlated with log$L_{\textrm{X}}$, while the surface gravity $\log g$ is negatively correlated with log$L_{\textrm{X}}$. The latter two correlations are somewhat weaker, but still at the $ >3\sigma$ level. Regarding the X-ray activity level log($L_{\textrm{X}}$/$L_{\textrm{bol}}$), it has a negative correlation with the temperature $\log T_{\rm eff}$ ($>5\sigma$), while being almost independent of [Fe/H] and $\log g$. 

\par

\subsection{Binary components parameters analysis}\label{sec:binaryparameters}

\par

For the SRASS sample, we collected 23 sources (hereafter, sub-SRASS) that feature measurements of their absolute stellar parameters for each of the two individual components, that is: mass ($M_{1}$ and $M_{2}$), radius ($R_{1}$ and $R_{2}$), and temperature ($T_{1}$ and $T_{2}$), as well as the orbital inclination ($i$) from the literature. Here, the more massive component of the binary is defined as Star$_{1}$, and its physical parameters have the subscript of 1, while the parameters with the subscript of 2 represent the less massive component Star$_{2}$. These parameters are listed in Table~\ref{table:SRASSsamples}. The K-S tests between the sub-SRASS and the whole SRASS samples are performed upon the orbital period and log$L_{\textrm{X}}$. The null-hypothesis probabilities, $P_{\rm K-S}$,  are 79.5\% and 37.9\%, respectively, suggesting that they follow the same distributions. For the STW sample, there are no such measurements for the absolute stellar parameters for each of the binary components. 

\par

 \begin{table*}[!htpb]
   \caption{Absolute stellar parameters for each component of the EWXs in sub-SRASS}
   \begin{center}
   \begin{tabular}{lcccccccccccc}\hline \hline

Name &  log$L_{\textrm{X}}$ & Period & $M_{1}$ & $M_{2}$  &  $R_{1}$ & $R_{2}$  & $T_{1}$  & $T_{2}$ & $i$ &  Subtype &  Reference \\
 & (erg s$^{-1}$) & (days) & (M$_{\odot}$) & (M$_{\odot}$) &  (R$_{\odot}$) & (R$_{\odot}$) & (K) & (K) & ($^{\circ}$) & &  \\ \hline
V523 Cas & 29.707 & 0.2337 & 0.740 & 0.380 & 0.770 & 0.590 & 5104 & 5076 & 84.36  &  A & (1) \\
BX Peg   & 29.880 & 0.2804 & 1.020 & 0.380 & 0.966 & 0.623 & 5872 & 5300 & 87.00  &  A & (2) \\
SX Crv   & 30.376 & 0.3166 & 1.246 & 0.098 & 1.347 & 0.409 & 6340 & 6160 & 61.21  &  A & (3) \\
V508 Oph & 29.983 & 0.3448 & 1.010 & 0.520 & 1.060 & 0.800 & 5980 & 5893 & 83.78  &  A & (4) \\
GR Vir   & 30.081 & 0.3470 & 1.370 & 0.170 & 1.420 & 0.610 & 6300 & 6163 & 83.36  &  A & (5) \\
AH Cnc   & 30.488 & 0.3604 & 1.188 & 0.185 & 1.332 & 0.592 & 6300 & 6151 & 83.11  &  A & (6) \\
U Peg    & 30.161 & 0.3748 & 1.149 & 0.379 & 1.224 & 0.744 & 5860 & 5785 & 77.51  &  A & (7) \\
V566 Oph & 29.859 & 0.4096 & 1.500 & 0.380 & 1.490 & 0.810 & 6456 & 6247 & 80.40  &  A & (8) \\
AQ Psc   & 30.470 & 0.4756 & 1.260 & 0.280 & 1.220 & 1.180 & 6445 & 5946 & 68.90  &  A & (9) \\
SW Lac   & 30.453 & 0.3207 & 1.207 & 0.991 & 1.090 & 1.000 & 5371 & 5529 & 80.95  &  W & (10)\\
RW Com   & 29.638 & 0.2373 & 0.800 & 0.380 & 0.770 & 0.540 & 4720 & 4900 & 74.90  &  W & (11)\\
RW Dor   & 29.919 & 0.2854 & 0.820 & 0.520 & 0.881 & 0.703 & 5238 & 5560 & 76.90  &  W & (12)\\
BW Dra   & 30.198 & 0.2923 & 0.920 & 0.260 & 0.980 & 0.550 & 5980 & 6164 & 74.42  &  W & (13)\\
TW Cet   & 30.241 & 0.3117 & 1.060 & 0.610 & 0.990 & 0.760 & 5450 & 5600 & 83.70  &  W & (14)\\
FG Hya   & 30.139 & 0.3278 & 1.444 & 0.161 & 1.405 & 0.591 & 5900 & 6012 & 82.25  &  W & (15)\\
V781 Tau & 29.993 & 0.3449 & 1.060 & 0.430 & 1.130 & 0.760 & 5536 & 6000 & 65.89  &  W & (16)\\
AC Boo   & 30.060 & 0.3524 & 1.200 & 0.360 & 1.190 & 0.690 & 6241 & 6250 & 86.03  &  W & (17)\\
V752 Cen & 30.285 & 0.3702 & 1.310 & 0.390 & 1.300 & 0.770 & 6014 & 6138 & 82.07  &  W & (18)\\
RT LMi   & 29.662 & 0.3749 & 1.290 & 0.490 & 1.280 & 0.840 & 6400 & 6513 & 84.10  &  W & (19)\\
TX Cnc   & 30.182 & 0.3829 & 1.350 & 0.610 & 1.270 & 0.890 & 6250 & 6537 & 62.10  &  W & (20)\\
UV Lyn   & 30.083 & 0.4150 & 1.430 & 0.550 & 1.400 & 0.920 & 5736 & 5960 & 66.13  &  W & (16)\\
UX Eri   & 30.993 & 0.4453 & 1.450 & 0.540 & 1.450 & 0.910 & 6046 & 6100 & 76.89  &  W & (21)\\
V502 Oph & 30.263 & 0.4534 & 1.370 & 0.460 & 1.510 & 0.940 & 5900 & 6140 & 76.40  &  W & (22)\\

\hline\noalign{\smallskip}
  \end{tabular}
  \end{center}
  \label{table:SRASSsamples}
    \begin{tablenotes}
     \footnotesize
    \item[1] Ref.(1):\citet{2016NewA...44...78M}, (2):\citet {2015NewA...41...17L}, (3): \citet {2004AcA....54..299Z}, (4): \citet{2015AJ....149...62X} , (5): \citet{2004AJ....128.2430Q} , (6): \citet{2016RAA....16..157P} , (7): \citet{2002CoSka..32...79P} , (8): \citet{2018Ap&SS.363...34S} , (9): \citet{2020MNRAS.491.6065Z} ,  (10): \citet{2014arXiv1402.2929E} , (11): \citet{2011A&A...525A..66D} , (12): \citet{2019PASJ...71...34S} , (13): \citet{1986AJ.....92..666K} , (14): \citet{1982A&AS...47..211R} , (15): \citet{2005MNRAS.356..765Q} , (16): \citet{2020ApJ...901..169L} , (17): \citet{2010IBVS.5951....1N}, (18): \citet{2019MNRAS.489.4760Z} , (19): \citet{2008PASJ...60...77Q} , (20): \citet{1987PNAS...84..610C},  (21): \citet{2007AJ....134.1769Q} , (22): \citet{2016PASJ...68..102X}
    \end{tablenotes}
\end{table*}

\par

\begin{figure*}
\centering
\includegraphics[width = 14cm]{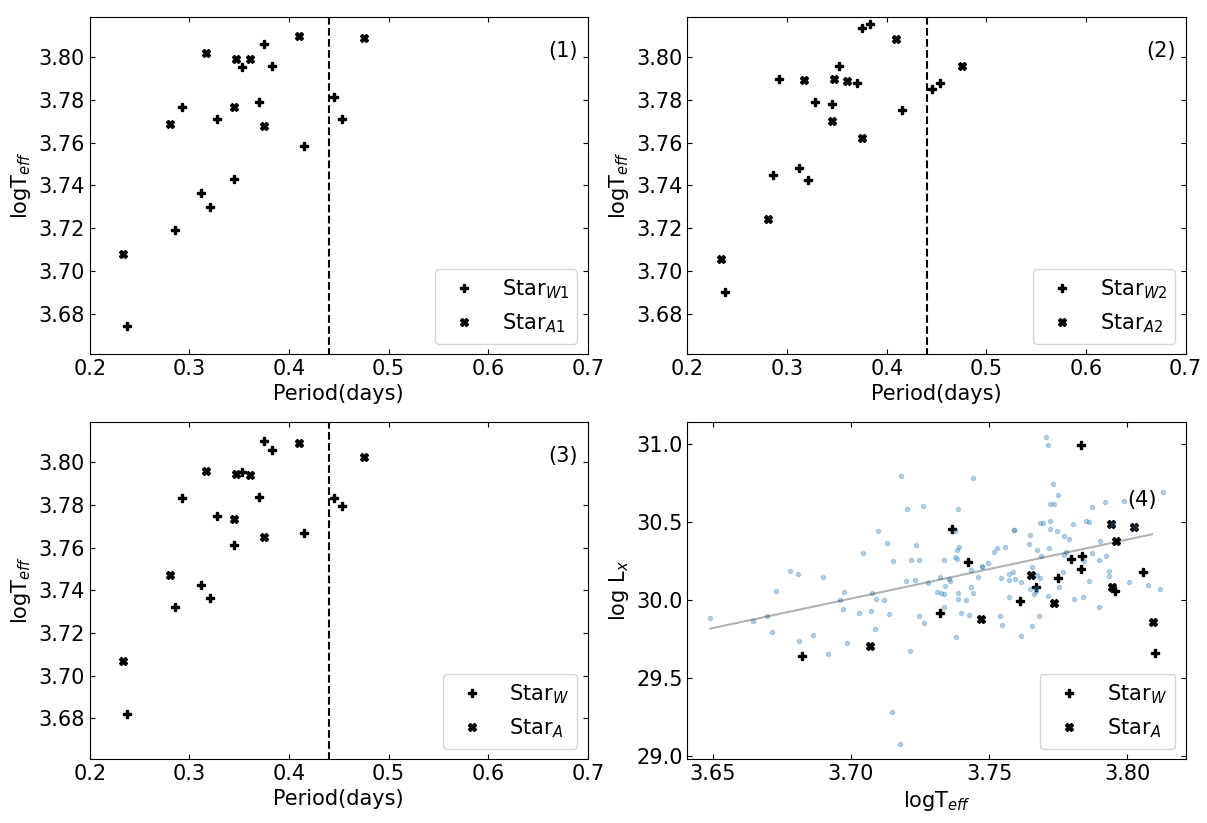}
\caption{Plots of the period vs. log$T$ for primary stars (1), secondary stars (2), and the binary systems (3) of the W-subtype (`+' symbols) and A-subtype (`x' symbols) from sub-SRASS. 
 (4): Plot of log T vs. log$L_{\textrm{X}}$. The dashed lines are at $P=0.44$ days. The light blue points and gray line in the background of panel (4) are the Spec-EWX sample points and its fitting line from Figure~\ref{fig:LAMOST_logT_logg_FeH_Lx_N} (1-2).}
\label{fig:DIS_T_Lx}
\end{figure*}

\begin{figure*}
\centering
\includegraphics[width = 14cm]{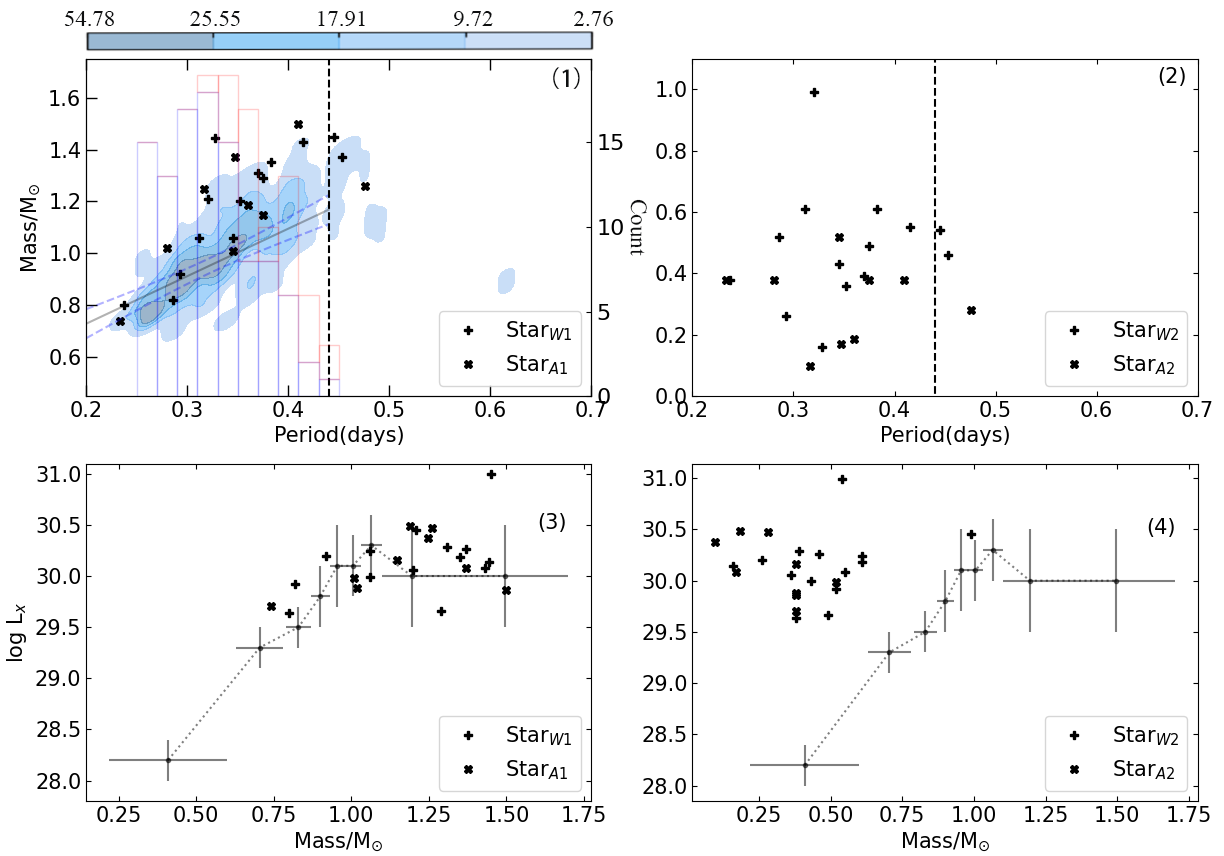}
\caption{Plots of the period vs. stellar mass for primary (1) and secondary (2) from sub-SRASS, and the mass of primary (3) and secondary (4) vs. log$L_{\textrm{X}}$. The black dashed line in panels (1) \& (2) is at 0.44 days. The small black points with dashed line in panels (3) \& (4) indicate the X-ray luminosity of the main sequence star at saturation in the mass range (0.22 to 1.7 M$_{\odot}$). In panel (1), the blue contours contain the data points from the 10,000 simulated samples (see text) with masses corresponding to the spectral temperatures from Spec-EWX (extra adding $0$--200~K random errors of uniform distribution for each object) against the period, while the gray line and  light blue dashed lines are the best-fit relation of these points with periods of $\leq$~0.44~days and its 95\% uncertainty range. The blue and red histograms (with 0.02-day steps) with axis on the right represent the period distributions of the primary stars with masses less than 1.1~M$_{\odot}$ and greater than 1.1~M$_{\odot}$, respectively (the average value of 10,000 simulations). }
\label{fig:DIS_M_Lx}
\end{figure*}

We investigate the possible relationships between log$L_{\textrm{X}}$ and the temperature and mass of the different components in Figures~\ref{fig:DIS_T_Lx} and \ref{fig:DIS_M_Lx}, respectively. The temperatures of the primary (more massive), secondary, and the binary (average value of the two components) are plotted in Figure~\ref{fig:DIS_T_Lx} (1), (2) and (3), respectively. The primary and secondary components, as well as the W/A-subtypes, are indicated in the subscripts of the labels in each panel. The black dashed lines are located at $P=0.44$ days, the break point determined in Section~\ref{sec:X-ray_P}. For all three panels, the temperature increases along with the period when $P < 0.44$ days. The similarity in this tendency is expected since the temperature difference between the two components of an EW binary is small, about several hundred kelvins. The relation between the average temperature of two components and X-ray luminosity is also plotted in panel (4). We also plot Spec-EWX sample and its best-fit line (see Figure~\ref{fig:LAMOST_logT_logg_FeH_Lx_N}, panel 1-2) in light blue points and gray line in the background, respectively. In this panel, the sub-SRASS objects are distributed in the region generally similar to where the Spec-EWX sample occupies. 

\par

Similarly, we plot the masses of the primary and secondary from sub-SRASS in Figure~\ref{fig:DIS_M_Lx} (1) and (2), respectively. The relation between log$L_{\textrm{X}}$ and the masses of primary and secondary are plotted in panels (3) and (4). In the last two panels, the black dots with error bars represent the typical saturated X-ray luminosity of the main sequence stars within the mass range (0.22 to 1.29 M$_{\odot}$). Each horizontal error bar indicates that stars with that mass range reach saturation when $P < 1$ day at the corresponding X-ray luminosity  \citep[see][Table~3]{2003A&A...397..147P}. 
In addition, for those EWXs with the mass of the primary star between $\sim$1.29 and $\sim$ 1.7 M$_{\odot}$ (corresponding to F-type), we adopted a saturation X-ray luminosity value of $\sim$3.2 $\times$ 10$^{29}$ to $\sim$3.2 $\times$ 10$^{30}$ erg s$^{-1}$ \citep{2019A&A...628A..41P, 2020ApJ...902..114W}.

\par

The mass ranges of the primary and secondary stars are  $\sim$0.7 to $\sim$1.5 M$_{\odot}$ and $\sim$ 0.1 to $\sim$1 M$_{\odot}$, respectively. It is clearly shown in panels (1) that the mass of the primary star is positively correlated with the period below $P=0.44$ days, while the mass of the secondary shows a much weaker correlation (panel 2). For all the relationships discussed in this subsection, it is clear that there is no significant difference between the behaviors of W-subtype and A-subtype systems.

\par

\par

\par

\section{Discussion}\label{sec:Discu}

\subsection{Sample completeness}\label{sec:completeness}

In this section, we discuss the completeness of our EWX sample, especially with regard to how the X-ray nondetections would affect our results. For our reliably classified EW binaries without X-ray detections, most of the X-ray luminosity upper limits have been obtained from the XMM Slew Survey or the RASS (see the gray symbols in the upper two panels of Figure~\ref{fig:Lx_P}). These upper limits are generally substantially higher than the X-ray luminosity range of detected EWXs because of the low sensitivity of the two surveys. For the correlation analyses we  carried out (presented in Section~\ref{sec:data_analysis}), adding these upper limits will not provide further constraints.

From the pointed XMM-Newton observations that were utilized in the 4XMM-DR9 catalog, we obtained X-ray luminosity upper limits for 39 EWXs. These objects and the 255 EWXs detected by XMM-Newton in Group A are all from the 14096 EW binaries that we selected using the same criteria as those of the Group A (after applying the cuts on period, distance, and parallax precision; see Section~\ref{sec:sec:upper limits}). Therefore, the covering fraction (the number of EWXs over the number of EWs) of pointed XMM-Newton observations for EWXs in the whole sky region is estimated to be $2.09\%$, which is similar to the $2.85\%$ sky coverage of XMM-Newton pointed observations (1152 deg$^{2}$).

We repeated the correlation analysis between the period and X-ray luminosity by taking those 39 upper limits into account. We employed the software package ASURV Rev 1.2 \citep{1990BAAS...22..917I, 1992ASPC...25..245L}, which implements the methodology proposed by \citet{1986ApJ...306..490I} and contains the Expectation-Maximization regression algorithm for astronomical data with detection limits, that is, censored data. The best-fit correlations of $P$-$\rm log L_{\rm X}$, $P$-$\rm log L_{\rm bol}$, and $P$-$L_{\textrm{X}}$/$L_{\textrm{bol}}$ for the objects with periods of less than 0.44~days are $\log L_{\rm X} = 2.84(27) \times P + 29.07(9)$, $\log L_{\rm bol} = 4.34(31) \times P + 32.38(10),$ and $\log L_{\textrm{X}}/L_{\textrm{bol}}= -1.42(29) \times P -3.37(10)$, which are values that are highly consistent with previous results on X-ray detections only shown in Equation~(\ref{equ:STW_LX_P}) and Table~\ref{table:ste_sub_srassLX/Lbol}. This indicates that whether or not including the upper limits from pointed XMM-Newton observations would not affect the correlation studies of EWXs. We do not include these upper limits in the discussions in Sections~\ref{sec:magnetic}--\ref{sec:primary_Xemission}.

There are four objects having tight X-ray upper limits from the pointed XMM-Newton observations that are well below the best-fit period-luminosity relation of EWXs. Labeled as bigger magenta `$tri\_down$' symbols in Figure~\ref{fig:Lx_P}~(1) from left to right, these objects are ASASSN-V J152949.56-445113.1, J023156.36+605519.0,  J074108.82+251600.7, and J071826.31-243845.6. 
The exposure times of their corresponding observations are 14ks, 7ks, 9ks, and 8ks, respectively, suggesting that they do not suffer from insufficient exposure depth. They do indeed have lower X-ray luminosity than typical EWXs. The estimated temperatures from Gaia DR2 for these four objects (4198K, 4864K, 6655K, and 6948K) may provide clues to the reason. Based on the temperature and period information from Spec-EWX sample in Table~\ref{table:Spectra_EWXs}, we calculate the average temperature of the six objects with closest periods to each of the four sources. The mean temperature values are 5500 $\pm$ 77K, 5831 $\pm$ 84K, 6113 $\pm$ 102K, and 6098 $\pm$ 61K, respectively. The temperatures of the four sources are $> 3\sigma$ below or above their respective mean temperature. This may indicate that the temperature deviation of EW-type binaries may affect their X-ray emission. As discussed in Section~\ref{subsubsec:Lx_Lbol_P} (below), EW-type binaries with lower temperature may generate less overall radiation across the full spectrum, while stars hotter than typical would also produce weaker X-ray radiation due to their thinner convection zone. Deeper X-ray observations on these sources are needed to replace the current upper limits with definite X-ray luminosity level and, thus, to facilitate further investigations on their lower X-ray emission level compared to typical EW binaries.

\subsection{$L_{\rm X}$ and $L_{\textrm{X}}$/$L_{\textrm{bol}}$ of whole EWX system}\label{sec:magnetic}

\subsubsection{Linear relationships with $P$}\label{subsubsec:Lx_Lbol_P}

\par

As shown in Figures~\ref{fig:Lx_P} and \ref{fig:Lx_Lbol_STW_sub_SRASS}, we obtained clear correlations ($>5\sigma$) of $P$-log$L_{\textrm{X}}$ and $P$-log($L_{\textrm{X}}$/$L_{\textrm{bol}}$) for EWXs with $P<0.44$~days using the high-quality X-ray and optical data of 4XMM-DR9 and Gaia DR2, and we provide the first linear parametrization of these relationships in this work. The same correlations for the SRASS sample are generally consistent ($\simlt 1\sigma$) with those of the STW sample (see Equations~\ref{equ:STW_LX_P} \& \ref{equ:SRASS_LX_P} and Table~\ref{table:ste_sub_srassLX/Lbol}). Since the STW has a larger sample size and higher-quality X-ray data from XMM-Newton, the  analyses presented in the following are mainly based on the correlation analysis results of the STW objects with periods of less than 0.44~days. As the number of EWXs with $P >0.44$~days is very small, only a qualitative description can be given, which is that the above relationships may appear to remain flat or weakly negative. 

\par

Based on the best-fit linear relationships we have obtained, the period can be treated as a good predictor of the X-ray luminosity (Section \ref{sec:X-ray_P}) and activity level for EWXs. The slope of the $P$-$\log(L_{\textrm{X}}$/$L_{\textrm{bol}}$) relation is $-1.36\pm0.29$, which is fully consistent with the difference between the slopes of the $P$-$\log L_{\textrm{X}}$ and $P$-$\log L_{\textrm{bol}}$ relationships ($2.81\pm0.27$ and $4.19\pm0.21,$ respectively). When the period of an EWX is shorter than 0.44~days, both its X-ray luminosity and bolometric luminosity rise with the increasing period, but the growth rate of former is slower than the latter, making the $P$-log($L_{\textrm{X}}$/$L_{\textrm{bol}}$) relation display a downward trend. The linear relationships of $P$-log$L_{\textrm{X}}$, $P$-log$L_{\textrm{bol}}$, and $P$-log($L_{\textrm{X}}$/$L_{\textrm{bol}}$), as listed in Equations~(\ref{equ:STW_LX_P})--(\ref{equ:SRASS_LX_logP}), and Table~\ref{table:ste_sub_srassLX/Lbol}, provide a convenient check of whether a given EWX has X-ray luminosity, bolometric luminosity, and X-ray activity level that are consistent with the typical population.

\par

With the period increasing from 0.2 to 0.44 days, the average X-ray luminosity, $L_{\rm X}$, of EWXs increases from 4.60$\times 10^{29}$ erg s$^{-1}$ to 2.17$\times 10^{30}$ erg s$^{-1}$, while the average bolometric luminosity, $L_{\rm bol}$, rises from 2.03$\times 10^{33}$ erg s$^{-1}$ to 2.04$\times 10^{34}$ erg s$^{-1}$. This makes that the average ratio of the X-ray radiation flux to the total radiation flux $L_{\rm X}/L_{\rm bol}$ decreases from $2.2 \times 10^{-4}$ to $1.0 \times 10^{-4}$. If we assume that the X-ray emission region is proportional to the total surface region of the EWXs, one reason for this phenomenon may be that the EWXs with short periods have higher X-ray activity level due to their thicker convective zone on its surface. Given the parallel correlation of increasing effective temperature with increasing period (i.e., $P$-$\log L_{\textrm{bol}}$), as the temperature rises, the convection zone on the surface will be thinner to generate less X-ray per unit area, while convection and rotation are the essential ingredients to power the magnetic dynamos \citep{1966ApJ...144..695W, 1967ApJ...150..551K, 2003A&A...397..147P}. Therefore, the X-ray emitting area of EWXs with shorter periods is smaller than that of EWXs with longer period, which makes the total X-ray luminosity increase along with the period. However, the convection zone of EWXs with short periods is thicker, providing higher X-ray activity level than that of longer-period EWXs.

\par

\subsubsection{Plateau of saturation}\label{subsubsec:Lx_Lbol_platreau}

\par

As shown in Figure~\ref{fig:LAMOST_logT_logg_FeH_Lx_N}~(1-3), for our Spec-EWX sample, the log($L_{\textrm{X}}$/$L_{\textrm{bol}}$) shows a monotonically decreasing trend with the increasing temperature, that is, the higher the temperature, the lower the X-ray activity level. This finding supports the magnetic dynamo theory we discuss in Section~\ref{subsubsec:Lx_Lbol_P}. However, based on the $B-V$ color index, \citet{2001A&A...370..157S} divided EWXs into cool ($>$ 0.6) and hotter ($<$ 0.6) stars, and concluded that the log($L_{\textrm{X}}$/$L_{\textrm{bol}}$) (also called normalized X-ray flux) of cool variables reaches a plateau while that of hotter objects decreases with the decreasing color index.
This means that at $B-V=0.6$ ($T_{\textrm{eff}}~\approx~5880$~K and log$T_{\textrm{eff}}\approx~3.77$), there is a "break" in the log$T_{\textrm{eff}}$-log($L_{\textrm{X}}$/$L_{\textrm{bol}}$) relation.\footnote{One should note that the $B-V$ color index corresponds almost linearly with temperatures between 0.30 and 1.15 ($T=$ 7300~K and 4410~K, respectively; \citealt{2000asqu.book.....C}).} We suggest that the difference between our results and that of \citet{2001A&A...370..157S} is due to the data completeness on the cool end. 
\citet{2001A&A...370..157S} stressed that it was implausible to determine the positive correlation between the X-ray activity level and color index with their available data  \citep[see][Section 3.4]{2001A&A...370..157S}.

\par

Similarly, the STW objects also show a negative $P$-log($L_{\textrm{X}}$/$L_{\textrm{bol}}$) correlation, namely,  objects with shorter periods generally have higher X-ray activity level, as shown in Figure~\ref{fig:Lx_Lbol_STW_sub_SRASS} (2). The nature of this similarity is that EWXs has a strong period-temperature (or bolometric luminosity) correlation. \citet{2001A&A...370..157S} invoked a surface horizontal flow to explain the X-ray emission level of EWXs.
As the period gradually reaches $\sim$0.2~days, the activity level of EWXs gradually approaches the state of saturation limit (log($L_{\textrm{X}}$/$L_{\textrm{bol}})=-3$), but there is no saturation plateau similar to that of ultra fast rotating stars (UFRs). We further discuss the interpretation of this relation in the last paragraph of Section~\ref{sec:primary_Xemission}.

\par

At $P<$0.44~days, the monotonously decreasing trend of the X-ray activity level (i.e., no plateau) may indicate that the EWXs are all in the same state (i.e., regarding whether they are in X-ray saturation or supersaturation). In the case of a mixture of multiple states, it is likely that the correlation will exhibit a changing slope or no correlation exists at all. For late-type main-sequence stars, the X-ray radiation reaches saturation when the rotation period is less than one day \citep{1984A&A...133..117V, 1987ApJ...321..958V, 1993ApJ...410..387F, 2003A&A...397..147P}. For EWs, the orbital period is synchronized with the rotational period of their individual components. These EWX objects have a period far less than one day, which suggests that their X-ray emission is likely to reach saturation as well. 

\par

\subsection{Binary components versus X-ray emission}\label{sec:primary-dominating}

\par

In this section, we compare the X-ray luminosity of the EWXs with that of the main sequence stars for given mass ranges when they reach saturation, as shown by the black points with error bars in Figure~\ref{fig:DIS_M_Lx} (3) and (4). For the main sequence stars, the saturated X-ray luminosity gradually increases from $\sim$ $1.2\times10^{29}$ erg s$^{-1}$ to $\sim$4.0 $\times$ 10$^{30}$ erg s$^{-1}$ with increasing stellar mass from 0.6 to 1.1 M$_{\odot}$.  However, as the mass of the stars continues to grow (in the range of 1.1 to 1.7 M$_{\odot}$), the saturated X-ray luminosity remains at $\sim10^{30}$ erg s$^{-1}$. The panel (3) demonstrates that in the range of $\sim$0.6 to $\sim$1.5 M$_{\odot}$, the X-ray luminosity of an EWX, whether for A-subtype or W-subtype, is generally consistent with the saturated X-ray luminosity of a single main sequence star with the same mass value as that of the primary component of this EWX, but not the mass of the secondary (panel 4). More generally, we further infer that EWXs may inherit the changing trends of the X-ray luminosities of the single stars with different masses. These may be the results of a combination of the primary star dominating the X-ray radiation of the binary system \citep{2004A&A...426.1035G, 2006ApJ...650.1119H} and the X-ray emission being saturated for the EWXs with $P<$0.44~days.

\par

Given that the X-ray emissions of single stars and EWXs are all determined by the coverage area of the X-ray emitting regions on their surface \citep{2001A&A...370..157S}, the consistent X-ray luminosities of EWXs and single stars we mention above may imply that the coverage area of the EWXs is close to that of the single star with the mass of the primary component. In this scenario, it can be understood that the different log($L_{\textrm{X}}$/$L_{\textrm{bol}}$) distributions exhibited by UFRs and EWXs with the same period \citep{2001A&A...370..157S} is mainly related to the differences in the bolometric flux of UFRs and EWXs. The bolometric luminosity of an EWX binary is likely greater than that of an UFR (e.g., those from \citealt{2003A&A...397..147P}) with similar mass to the primary star of that EWX, which will result in lower log($L_{\textrm{X}}$/$L_{\textrm{bol}}$) of EWXs. Meanwhile, the difference between the bolometric luminosity of UFRs and EWXs also likely varies with changing periods (for $P<$~0.44 days). For EWXs, the $P$-$L_{\textrm{bol}}$ relationship (Figure~\ref{fig:Lx_Lbol_STW_sub_SRASS} panel 1) is essentially the same relationship between the period and mass (Section~\ref{sec:primary_Xemission}). Based on this degeneracy, we can find that the $P$-log($L_{\textrm{X}}$/$L_{\textrm{bol}}$) trend of our EWXs (all with primary mass greater than 0.5M$_\odot$) is basically similar to the mass-log($L_{\textrm{X}}$/$L_{\textrm{bol}}$) relation of single stars under saturation  in the middle panel of Figure 7 (we note the direction of the mass axis in that figure) in \citet{2003A&A...397..147P}, which supports our above analyses.

\par

\subsection{Period and mass versus X-ray emission}\label{sec:primary_Xemission}

\par

Since the LAMOST DR7 survey used single-star models to fit unresolved binaries, \citet{2018MNRAS.473.5043E} derived that the LAMOST temperatures derivation is systematically lower than the temperature from the spectrum of the primary star by  $\simlt 200$~K. Therefore, we generated 10,000 simulated samples by adding random errors uniformly distributed in [0, 200]~K to the spectral temperature of each Spec-EWX object. Then we derived their corresponding masses 
(linear interpolated using Tables 15.7 and 15.8 of \citealt{2000asqu.book.....C}) and performed linear fittings between the mass and period of the simulated samples with $P<$~0.44~days. The blue-colored contours in Figure \ref{fig:DIS_M_Lx} (1) contain the 10,000 simulated samples, while the gray line and light blue dashed lines represent the best-fitting relations between period and mass, and its 95\% uncertainty range, respectively.\footnote{This blue bubble structure in the lower right corner is due to the simulation of one binary system ASASSN-V J141756.07+210554.7, which has a period of 0.61 days but a spectral temperature of only 5321K.}  As shown in this figure, the primary stars' masses of sub-SRASS are mostly distributed in the range of that of Spec-EWX. We find that $83.7\pm$2.7~\% (from statistical distribution and error of the 10,000 simulated samples) objects of EWXs have masses lower than 1.1~M$_{\odot}$ in this sample. The best-fit gray line (slope is 1.82~$\pm$~0.02, intercept is 0.35~$\pm$~0.04) is intercepted by the black dashed line corresponding to $P=0.44$~days at mass value of $1.15\pm$0.04~M$_{\odot}$. Combined with the fact that the X-ray luminosity increases with mass until reaching $\sim$~1.1~M$_{\odot}$ (see Figure~\ref{fig:DIS_M_Lx}, panel 3) and the X-ray emission of EWXs is dominated by the primary, this naturally explains the positive correlation between period and $\log L_{\rm X}$ until $P=0.44$~days (corresponding to $\sim1.1$M$_{\odot}$) that we found in Section~\ref{sec:X-ray_P}. Considering each finer period interval (see the blue and red histograms with axis on the right in Figure~\ref{fig:DIS_M_Lx} panel 1), the fraction of objects with a primary mass lower than 1.1~M$_{\odot}$ generally decreases with increasing period. It is apparent that when the period approaching $0.44$~days, the objects with primary more massive than 1.1~M$_{\odot}$ start to dominate. Meanwhile, the X-ray luminosity of the primary stars in this mass range ($>1.1$~M$_{\odot}$) remains constant ($\sim$ 10$^{30}$ erg s$^{-1}$). Therefore, the $P-\log L_{\rm X}$ correlation breaks at $P=0.44$~days.

\par

Certainly, we cannot rule out the contribution of secondary stars to the binary X-ray luminosity. Long-period binaries also usually have larger secondary stars. This would reinforce the positive correlation between X-ray luminosity and period. On the other hand, this would not affect the correlation break. Based on the mass range of the secondary, their saturated X-ray luminosity is $\geqslant 1.5$ dex lower than that of the primary. 
The X-ray emission from the primary remains dominating. It is notable that the sub-SRASS sample has a higher fraction of objects with more massive primary ($>1.1$~M$_{\odot}$) than that of the Spec-EWX sample, as shown in Figure~\ref{fig:DIS_M_Lx} (1). This could be a selection effect since obtaining the spectral parameters for each of the binary component would require higher quality optical spectra from more luminous (thus more massive) stars.

\par

As shown in Figure~\ref{fig:DIS_M_Lx} (1), the increase of the main mass is strongly correlated with the increase of the period of EWXs. Given the positive correlation between effective temperature and mass, at a certain mass, the temperature of primary star is so high that the convective zone on its surface would not have adequate material to support a normal magnetic dynamo, which is essentially driven by convection and rotation \citep{1966ApJ...144..695W, 1967ApJ...150..551K, 2003A&A...397..147P}. This critical mass/temperature relation occurs at the mass of a primary star is $\geqslant 1.1$~M$_{\odot}$ (P$\sim$0.44 days), which may result in decreased X-ray luminosity for binaries with longer rotation periods. Most of EWXs in this work have short periods, which provide a unique sample to test the magnetic dynamo theories under the extreme physical conditions from fast-rotating stars.

\par

In combination with sections 4.1, 4.2, and 4.3, we can conclude that the physical nature of the $P$-log($L_{\textrm{X}}$/$L_{\textrm{bol}}$) relation for single stars and for EWXs is quite different. For single stars, the log($L_{\textrm{X}}$/$L_{\textrm{bol}}$) increases monotonically as the period decreases, meaning that they are not saturated with X-ray radiation. In a fixed mass range, their rotation periods are distributed over a wide range (e.g., 0.1 to 100 days). Only when the period is below a certain value will there be a plateau of X-ray luminosity and activity level \citep{2003A&A...397..147P, 2019A&A...628A..41P}, and this plateau represents the saturated X-ray activity level. Single stars with different masses have different saturated $P$-log$L_{\textrm{X}}$ and $P$-log($L_{\textrm{X}}$/$L_{\textrm{bol}}$) plateaus. For EWXs, as we discuss in Section \ref{subsubsec:Lx_Lbol_platreau}, although their $P$-log($L_{\textrm{X}}$/$L_{\textrm{bol}}$) relationship has a slope, they are still likely to be all saturated with X-ray emission. Given the primary star dominating the X-ray radiation of an EWX, the X-ray saturation luminosities of the EWXs are likely to behave as the collection of the X-ray saturation luminosities from single stars with a range of masses and, thus, $Mass$ and log$L_{\textrm{X}}$ are correlated with each other (Figure~\ref{fig:DIS_M_Lx} panel 3). Since the period and mass of EWXs are highly degenerated, the $P$-log$L_{\textrm{X}}$ relation in Figure~\ref{fig:Lx_P} (1) is similar to the trend of $Mass$-log$L_{\textrm{X}}$ in Figure~\ref{fig:DIS_M_Lx} (3), which explains why EWXs have no X-ray luminosity plateau although they are all in saturation. At $P<$0.44~days, $P$-log$L_{\textrm{X}}$ relation is the part that maintains a nearly monotonous increase. This, combined with the linear relationship of $P$-log$L_{\textrm{bol}}$, naturally leads to our finding that the $P$-log($L_{\textrm{X}}$/$L_{\textrm{bol}}$) relation maintains a monotony with no plateau. Essentially, for EWXs, the (primary) mass is probably the most fundamental physical parameter that determines the X-ray emission level, and the period and $L_{\textrm{bol}}$ are the observed manifestations of the mass. More EWXs with measured stellar parameters (such as mass, temperature of each component) and more single star X-ray data with wider mass ranges will strongly facilitate the study of X-ray radiation mechanisms of binary and single stars.

\par

\section{Summary}\label{sec:Summary}

\par

In this work, by cross-matching the AVSD with 4XMM-Newton DR9 and the 2RXS catalogs, we compile the largest sample to date of X-ray emitting EW-type binaries with periods of less than 1 day and distance less than 1~kpc. We also added in the RASS-selected EWXs from literature \citep{2001A&A...370..157S,2006AJ....131..633G,2006AJ....131..990C} and updated their distance and X-ray luminosity with the Gaia DR2 parallax data. The full EWX sample in this work contains 376 objects detected by XMM-Newton and 319 objects detected by ROSAT. For EW binaries with X-ray coverage but without any detections, most of the upper limits are from the low-sensitivity RASS and XMM slew survey, which could not provide useful constraints on the X-ray emission. There are 39 objects having tight X-ray upper limits from pointed XMM-Newton observations. Adding these upper limits to the sample does not affect the results of the correlation analysis. Our sample of EWXs is sufficiently complete for studying the X-ray properties of EW binaries (see Section~\ref{sec:completeness}).

The statistical distributions of parameters (period, temperature, mass, etc.) of EWXs and their relationships with the X-ray luminosity and X-ray activity level are investigated in Section~\ref{sec:data_analysis}. We discuss the properties of $\log L_{\textrm{X}}$, $\log L_{\textrm{bol}}$, and log($L_{\textrm{X}}$/$L_{\textrm{bol}}$) for the whole EWX systems in Section~\ref{sec:magnetic}. Then, at the level of component stars, we further investigate the possible physical mechanisms of the observed characteristics of EWXs in Sections~\ref{sec:primary-dominating} and \ref{sec:primary_Xemission}. Our main results are detailed as follows.

\begin{enumerate}

\par

\item  We provide the quantitative formulation of the strong positive correlation ($>5\sigma$) between the period, $P,$ and X-ray luminosity, log$L_{\textrm{X}}$, for EWXs (Equations~\ref{equ:STW_LX_P}--\ref{equ:SRASS_LX_logP}) at $P<0.44$~days. We also present the best-fit $P$-log$L_{\textrm{bol}}$ and $P$-log($L_{\textrm{X}}$/$L_{\textrm{bol}}$) relations for EWXs. These linear relationships effectively constrain the model of X-ray emission mechanisms for contact binary stars. Based on the fitting results of $P$-$\log L_{\textrm{X}}$ and $P$-$\log L_{\textrm{bol}}$, we provide the quantitative description of EWX activity levels with different periods, which may relate to the thickness of the convection zone in the magnetic dynamo theories. An EWX with a short period has a smaller but thicker X-ray emitting region than a relative longer period EWX. The relationship between log$T_{\rm eff}$-log($L_{\textrm{X}}$/$L_{\textrm{bol}}$) shows the higher the temperature, the lower the X-ray activity level, which is the evidence to support magnetic dynamo theories adopted to interpret the $P$-log($L_{\textrm{X}}$/$L_{\textrm{bol}}$) relation of EWXs.

\par

\item At $P<0.44$~days, neither the $P$-log($L_{\textrm{X}}$/$L_{\textrm{bol}}$) nor the log$T_{\rm eff}$-log($L_{\textrm{X}}$/$L_{\textrm{bol}}$) relationship has a saturation plateau similar to that of single ultra fast rotating stars. The X-ray activity level decreases monotonically with the period, which, combined with the short periods of EWXs, may indicate that EWXs are all in X-ray emission saturated state. 

\par

\item  We compiled the spectral parameters, including effective temperature, $T_{\rm eff}$,  metallicity [Fe/H], and surface gravity $\log g$, for the Spec-EWX sample using LAMOST spectroscopy. We find that EWXs and the general EW population have different statistical distributions ($>3\sigma$) of all the above three spectral parameters. In particular, the proportions of EWXs on the low mass (temperature) end is higher than that of EWs. The values of $T_{\rm eff}$ and [Fe/H] are positively correlated with the X-ray luminosity while $\log g$ is anti-correlated with log$L_{\textrm{X}}$. There is a high confidence ($>5\sigma$) negative correlation between temperature and activity level log($L_{\textrm{X}}$/$L_{\textrm{bol}}$).

\par

\item Based on the relation between saturated X-ray luminosity and the mass of main sequence stars, we find that the X-ray luminosity of an EWX is generally consistent with the saturated X-ray luminosity of a main sequence star with the same mass as the primary component of the EWX. The X-ray saturation luminosity values of the EWXs are similar to the collection of single stars with different masses when they all reach saturation.

\par

\item Most of our EWXs with $P < 0.44$~days have primary stars that are less massive than 1.1~M$_{\odot}$. In this mass range, the X-ray luminosity increases with mass while it remains constant when the primary mass exceeds 1.1~M$_{\odot}$. As the period increases, the temperature, primary mass and the X-ray luminosity of the binary systems change in concordance. The mass distribution of the primary stars may be the direct reason for the positive $P$-$\log L_{\textrm{X}}$ correlation and also contribute to break for this relation at $P\sim$ 0.44~days. Because there is no plateau in $P$-$\log L_{\textrm{X}}$ and the $P$-$\log L_{\textrm{bol}}$ remains monotonically increasing for periods less than 0.44 days, we  strictly confirm that there is a decreasing tendency in $P$-log($L_{\textrm{X}}$/$L_{\textrm{bol}}$) with no plateau. The degeneracy between the mass and the period of EWXs results in the monotonous relationships of $P$-log$L_{\textrm{X}}$, $P$-log$L_{\textrm{bol}}$ and $P$-log($L_{\textrm{X}}$/$L_{\textrm{bol}}$). 

\end{enumerate}

In conclusion, most, if not all, of the EWXs are in X-ray saturation state. The $P$-log($L_{\textrm{X}}$/$L_{\textrm{bol}}$) relation for EWXs is the manifestation of the X-ray saturation and the degeneracy between mass, period, and temperature. The mass of the primary star is the most fundamental physical parameter determining the X-ray emission properties of EWXs.

\begin{acknowledgements}

We thank the anonymous referee very much for the constructive suggestions, which helped to improve the paper. This work is supported by the National Natural Science Foundation of China (NSFC) under the grant numbers U1938105 (J.L. and J.Wu), U2031209 (A.E.), 11925301 (W.-M.G), 11973002 (M.Y.S), and U1831205 (J.Wang). We acknowledge the data support from Guoshoujing Telescope. Guoshoujing Telescope (the Large Sky Area Multi-Object Fiber Spectroscopic Telescope LAMOST) is a National Major Scientific Project built by the Chinese Academy of Sciences. Funding for the project has been provided by the National Development and Reform Commission. LAMOST is operated and managed by the National Astronomical Observatories, Chinese Academy of Sciences.
We also acknowledge the support of X-ray data based on observations obtained with \textit{XMM-Newton}, an ESA science mission with instruments and contributions directly funded by ESA Member States and NASA. This work has made use of data from the European Space Agency (ESA) mission \textit{Gaia} (https://www.cosmos.esa.int/gaia), processed by the \textit{Gaia} Data Processing and Analysis Consortium. We thank the Las Cumbres Observatory and its staff for its continuing support of the ASAS-SN project. ASAS-SN is supported by the Gordon and Betty Moore Foundation through grant GBMF5490 to the Ohio State University, and NSF grants AST-1515927 and AST-1908570. Development of ASAS-SN has been supported by NSF grant AST-0908816, the Mt. Cuba Astronomical Foundation, the Center for Cosmology and AstroParticle Physics at the Ohio State University, the Chinese Academy of Sciences South America Center for Astronomy (CAS-SACA), the Villum Foundation, and George Skestos.

Software: Matplotlib \citep{2007CSE.....9...90H}, Numpy (https://numpy.org/), Scipy (http://www.scipy.org), Seaborn \citep{2021JOSS....6.3021W}.

\end{acknowledgements}

\par

\end{document}